\documentclass[12pt]{article}
\usepackage[utf8]{inputenc}

\usepackage{adjustbox}
\usepackage{algorithm2e}
\usepackage{amsmath}
\usepackage{amssymb}
\usepackage{amsfonts}
\usepackage{mathtools}
\usepackage{multirow}
\usepackage{bbm}
\usepackage{bm}
\usepackage[authoryear]{natbib}
\usepackage{hyperref}
\hypersetup{colorlinks=true,linkcolor=blue,urlcolor=blue,citecolor=blue}
\usepackage{caption}
\usepackage{dsfont}
\usepackage{float}
\usepackage{graphicx}
\usepackage{subcaption}
\usepackage{authblk}

\newcommand\given[1][]{\:#1\vert\:}
\newcommand{\dd}{\mathop{}\!\mathrm{d}}

\title{Joint Models for Handling Non-Ignorable Missing Data using Bayesian Additive Regression Trees:\\Application to Leaf Photosynthetic Traits Data}
\author[1]{Yong Chen Goh}
\author[2]{Wuu Kuang Soh}
\author[3]{Andrew C. Parnell}
\author[1]{\newline Keefe Murphy}
\affil[1]{Hamilton Institute and Department of Mathematics and Statistics, Maynooth University}
\affil[2]{National Botanic Gardens of Ireland}
\affil[3]{School of Mathematics and Statistics, Insight Centre for Data Analytics, University College Dublin}
\date{}

\begin{document}

\def\spacingset#1{\renewcommand{\baselinestretch}%
{#1}\small\normalsize} \spacingset{1}

\maketitle

\begin{abstract}
Dealing with missing data poses significant challenges in predictive analysis, often leading to biased conclusions when oversimplified assumptions about the missing data process are made. In cases where the data are missing not at random (MNAR), jointly modeling the data and missing data indicators is essential. Motivated by a real data application with partially missing multivariate outcomes related to leaf photosynthetic traits and several environmental covariates, we propose two methods under a selection model framework for handling data with missingness in the response variables suitable for recovering various missingness mechanisms. Both approaches use a multivariate extension of Bayesian additive regression trees (BART) to flexibly model the outcomes. The first approach simultaneously uses a probit regression model to jointly model the missingness. In scenarios where the relationship between the missingness and the data is more complex or non-linear, we propose a second approach using a probit BART model to characterize the missing data process, thereby employing two BART models simultaneously. Both models also effectively handle ignorable covariate missingness. The efficacy of both models compared to existing missing data approaches is demonstrated through extensive simulations, in both univariate and multivariate settings, and through the aforementioned application to the leaf photosynthetic trait data.
\end{abstract}

\noindent%

\spacingset{1.9}

\section{Introduction}
\label{sec: 1 Introduction}
Missing data poses a challenge in predictive data analysis, persistently undermining the robustness and reliability of statistical modeling.  
This issue is exacerbated particularly in multivariate response settings, where missingness in one response may depend on the values of other responses which are also partially observed, and where missingness patterns can vary across cases.
In \citet{maire2015global}, the authors analyzed the \textit{global Amax} data to investigate the influence of 20 soil and 26 climate variables on 5 leaf photosynthetic traits: light-saturated photosynthetic rate (\textit{Aarea}), stomatal conductance (\textit{Gs}), leaf nitrogen (\textit{Narea}), leaf phosphorus (\textit{Parea}), and specific leaf area (\textit{SLA}). Despite having 46 fully observed covariates, the data exhibit substantial missingness in the responses, with only a small fraction of complete instances available for analysis. More specifically, out of $2368$ total cases, only $217$ were completely present. 
When confronted with partially-observed data such as this, analysts must navigate a choice between discarding missing samples, conducting imputations to fill in missing values for subsequent analysis, or employing tailored predictive models for incomplete data. Central to this decision-making process is the need to make reasonable assumptions about the underlying mechanisms governing the missingness probabilities --- specifically, whether the data are missing completely at random (MCAR), missing at random (MAR), or missing not at random (MNAR) \citep{rubin1976inference}.

Most methods for handling missing data rely on the assumptions of MCAR or MAR. Complete case analysis, a common approach that uses only fully observed rows, often results in significant information loss and biased results when missingness is not MCAR. Imputation strategies, such as mean/median imputation, multiple imputation \citep{van2018flexible}, and MICE \citep{azur2011multiple, van2011mice, little2019statistical}, use observed data to estimate missing values under MCAR or MAR assumptions. Methods like the Expectation-Maximization algorithm \citep[EM;][]{dempster1977maximum} and \texttt{missForest} \citep{stekhoven2012missforest}, a non-parametric technique using random forests, also address mixed-type data imputation.
Advanced predictive modeling methods, such as \texttt{BARTm} \citep{kapelner2015prediction} and \texttt{XGBoost} \citep{chen2016xgboost}, handle missing data in covariates without explicit imputation, although they are limited to univariate responses. \texttt{BARTm} incorporates missing covariates into tree-splitting rules, while \texttt{XGBoost} treats missing covariates as distinct categories during training.

Under MNAR missingness, handling missing data requires careful consideration and often involves more complex modeling techniques, as the probability of data being missing depends on the unobserved values themselves, i.e., the missingness is directly related to the information that is not captured in the data. Thus, ignoring the missingness would lead to information loss and severely biased conclusions.
If obtaining more data to reduce the level of MNAR missingness is infeasible, two of the more popular formulations of MNAR models that can be considered are, according to \citet{little2019statistical}, the selection model \citep{heckman1976common} and the pattern-mixture model \citep{glynn1986selection, little1993pattern}. 
These methods are distinguished by the factorization of the joint distribution $p(\mathbf{Y}, \mathbf{M})$, where $\mathbf{Y}$ is the partially-observed response variable and $\mathbf{M}$ is its missingness indicator. 

In the pattern-mixture model, the joint distribution is factorized into $p(\mathbf{Y},\mathbf{M})=p(\mathbf{M})p(\mathbf{Y}\given \mathbf{M})$.
The data are first stratified based on their missingness patterns, and then the response is modeled based on the specific missingness patterns.
This method is advantageous for investigating differences in the response distributions across distinct missing data patterns, and may be insightful for exploring the sensitivity of inferences to different missing mechanism assumptions \citep{michiels1999selection}.
However, pattern-mixture models often encounter under-identifiability issues, necessitating the imposition of model restrictions or the incorporation of prior information \citep{little1993pattern, thijs2002strategies, little2008selection}. 
In contrast, the selection model factorizes the joint distribution into $p(\mathbf{Y},\mathbf{M})=p(\mathbf{Y})p(\mathbf{M}\given \mathbf{Y})$, modeling the response as if no data were missing, followed by modeling the missingness probabilities given the response.
This approach directly estimates parameters related to the full population of interest and reflects the natural order of events where the response occurs before missingness is introduced \citep{little2008selection}.

In this paper, we focus on a novel selection model framework --- outlined later in Equation \eqref{eq: selection model factorization} --- to address the challenges posed by MNAR missing data in the context of predictive data analysis with multivariate outcomes. More specifically, we present two multivariate response predictive models arising from the selection model factorization to analyse the \textit{global Amax} data without restrictive assumptions on the missing data mechanism.
In both approaches, which we refer to as `missBART1' and `missBART2', we specify a multivariate Bayesian additive regression trees \citep[BART;][]{chipman2010bart, um2023bayesian, mcjames2024bayesian} model for the partially-observed responses. 
The key distinction between the two models lies in their approaches to modeling the missing data mechanisms.
 
In missBART1, we propose a multivariate probit regression model for the missingness. 
Probit regression was originally introduced in \citet{bliss1934method}; a full Bayesian model was later described in \citet{albert1993bayesian} and the multivariate extension was developed in \citet{chib1998analysis}. 
One benefit of the probit regression model is its parametric nature, which enables the characterization of different missing data mechanisms based on interpretable model parameters.
Through prior specifications within the probit regression model, we introduce additional flexibility by enabling the incorporation of prior beliefs regarding the underlying missing data mechanism, as well as allowing for efficient Gibbs sampling of the missing responses within the Bayesian framework.

In the probit regression model, the underlying latent structure is inherently linear, making it less suitable in cases where missingness depends on other variables in a non-linear fashion. 
To address this limitation, missBART2 adopts an alternative approach to modeling missingness via the specification of a multivariate probit BART model. 
This fully non-parametric joint model leverages BART's ability to capture complex, non-linear relationships within both the data and missingness sub-models.
Although missBART2 lacks the interpretable coefficients provided by the probit regression model in missBART1 and also requires a Metropolis-Hastings step to sample missing responses, BART’s variable selection feature mitigates the need for prior assumptions about the missing data mechanism. This is particularly advantageous when limited information on the missingness mechanism is available.
While both models are designed to handle missingness in the responses, they can also accommodate missing covariates, with the constraint that covariates are missing under the ignorability assumption.
Due to the parametric nature of probit regression, missBART1 requires prior imputation on the covariates.
For missBART2, the \texttt{BARTm} approach for handling missing covariates can be incorporated, thus obviating the need for covariate imputation.
A schematic diagram of both joint models is included in \ref{Appendix: schematic diagram}.

The remainder of this paper is structured as follows: Section \ref{sec: 2 selection model} outlines the selection model in the context of handling multivariate missing response data. 
Section \ref{sec: 3 Bayesian probit regression} describes the multivariate probit regression model within the Bayesian framework. 
Section \ref{sec: 4 BART} outlines the BART model under a multivariate framework, giving mathematical formulations and prior specifications, followed by its probit formulation. 
Section \ref{sec: 5 Joint models} explains our two novel models missBART1 and missBART2 in detail. 
In Section \ref{sec: 6 Simulation Study}, results from several simulation studies are discussed, along with comparisons made between other methods for handling missing data to showcase the benefits and importance of appropriately handling MNAR missing data via joint models. 
Examples with missingness in the covariates are also included.
In Section \ref{sec: 7 Real data example}, we describe the \textit{global Amax} data in detail and show results obtained from applying missBART1 and missBART2 to the data. 
The paper ends in Section \ref{sec: 8 Discussion} with concluding statements and discussions on limitations and future work.

\section{Selection Models for Non-Ignorable Missing Data}
\label{sec: 2 selection model}
We now give a brief outline of the selection model used for modeling partially-observed data such as the \textit{global Amax} data without restrictive ignorability assumptions.
We restrict the missingness to the multivariate response variables and assume that all covariates are fully observed, as is the case with the \textit{global Amax} data.
However, scenarios with ignorable covariate missingness are discussed in Sections \ref{sec: 5 Joint models} and \ref{sec: 6 Simulation Study}.

Given $n$ observations and $p$ responses, define the partially-observed responses $\Tilde{\mathbf{Y}}$ as an $n \times p$ matrix such that $\Tilde{Y}_{ij} = Y_{ij}^{obs}$ if $M_{ij}=1$ and $\Tilde{Y}_{ij} = Y_{ij}^{mis}$ otherwise, for $i=1,\ldots,n$, $j=1,\ldots,p$.
Here, $\mathbf{Y}^{obs}=\{Y_{ij}\colon M_{ij}=1\}$ refers to the set of observed responses, $\mathbf{Y}^{mis}=\{Y_{ij}\colon M_{ij}=0\}$ refers to the set of missing responses, and $\mathbf{X}$ denotes a fully observed set of covariates.
Under the selection model, the complete data likelihood is factorised as
\begin{equation} 
    p\left(\Tilde{\mathbf{Y}}, \mathbf{M}\given\mathbf{X}, \boldsymbol{\theta}, \boldsymbol{\psi}\right) = \underbrace{p\left(\Tilde{\mathbf{Y}}\given\mathbf{X}, \boldsymbol{\theta}\right)}_{\text{data model}} \times \underbrace{p\left(\mathbf{M}\given\mathbf{X}, \Tilde{\mathbf{Y}}, \boldsymbol{\psi}\right)}_{\text{missingness model}},
    \label{eq: selection model factorization}
\end{equation}
where $\boldsymbol{\theta}$ and $\boldsymbol{\psi}$ are sets of parameters in the data and missingness distributions, respectively. 
This selection model framework first came to prominence in \citet{heckman1976common}. 

The selection model commonly employs a `Heckit' specification, where the data model is a linear regression and the missingness model is a probit regression. For greater flexibility, missBART1 and missBART2 instead use multivariate BART \citep{um2023bayesian} for the data model.
Further details on multivariate BART are provided in Section \ref{sec: 4 BART}. In what follows, both models will assume that the set of observed responses $\mathbf{Y}^{obs}$ are fixed and known, while the set of missing responses $\mathbf{Y}^{mis}$ are to be estimated as part of the model updates.

missBART1 uses a multivariate probit regression model for the missingness model, incorporating efficient Gibbs sampling via parameter-expanded data augmentation \citep{talhouk2012efficient}. In contrast, missBART2 leverages a multivariate probit BART model, allowing it to capture non-linear relationships and complex interactions in the missingness mechanism. This provides greater flexibility compared to the parametric probit regression model in missBART1, which relies on interpretable coefficients. 
Both approaches enable the recovery of MCAR, MAR, and MNAR mechanisms, addressing the challenges of real-world missingness where assumptions about the missingness mechanism are often uncertain.

\section{Bayesian Probit Regression} \label{sec: 3 Bayesian probit regression}
The multivariate probit regression model \citep{chib1998analysis} generalizes the univariate probit model of \citet{albert1993bayesian} and assumes a correlated structure between multivariate binary outcomes.
Given a $p$-dimensional set of binary outcomes $\mathbf{M}_i = (M_{i1},\ldots,M_{ip})$, 
the well-known data augmentation scheme from univariate probit regression can be applied by introducing the $p$-dimensional latent variables $\mathbf{M}_1^\star,\ldots,\mathbf{M}_n^\star$ such that
\begin{align} \label{eq: mv probit latent}
    \mathbf{M}_i^\star &= \mathbf{B}^\top \mathbf{Z}_i + \boldsymbol{\epsilon}_i, \quad \boldsymbol{\epsilon}_i \overset{i.i.d.}{\sim}\mathcal{N}_p\left(\mathbf{0}, \mathbf{R}\right), \\
    M_{ij} &= \begin{cases}
        0 & \text{if } M_{ij}^\star \leq 0 \\
        1 & \text{if } M_{ij}^\star > 0.
    \end{cases} \nonumber
\end{align}
where $\mathcal{N}_p$ denotes the $p$-variate normal distribution, $\mathbf{B}$ is an $r \times p$ matrix of regression coefficients, $\mathbf{Z}_i$ is the set of $r$ predictors, and $\mathbf{R}$ is a $p \times p$ correlation matrix. 
Due to the challenging nature of specifying a prior distribution for correlation matrices, we use the parameter expansion strategy from \citet{talhouk2012efficient} to sample $\mathbf{B}$ and $\mathbf{R}$. 
A conjugate prior is assigned to $\mathbf{B}$ such that
\begin{equation}
    \mathbf{B}\given \mathbf{R} \sim \mathcal{MN}_{r \times p}\left(\mathbf{0}, \boldsymbol{\Psi}, \mathbf{R}\right),
    \label{eq: prior B}
\end{equation}
where $\mathcal{MN}_{r \times p}\left(\mathbf{0}, \boldsymbol{\Psi}, \mathbf{R}\right)$ is a matrix-normal distribution with an $r \times p$ mean matrix of zeros, $r \times r$ positive-definite matrix of dispersion hyperparameters $\boldsymbol{\Psi}$, and $p \times p$ positive-definite correlation matrix $\mathbf{R}$, which is in turn given the marginally uniform prior of \citet{barnard2000modeling}.
For more information on the sampling algorithm, see \citet{talhouk2012efficient}.

\section{Bayesian Additive Regression Trees} 
\label{sec: 4 BART}
BART is a Bayesian sum-of-trees regression model that has earned substantial recognition since its development due to its flexibility and robustness while making accurate probabilistic predictions. 
Since its development, BART has been extended in various ways to handle multivariate responses. 
\citet{um2023bayesian} developed a multivariate version of BART to handle multivariate skewed responses, \citet{mcjames2024bayesian} extended the Bayesian causal forests \citep[BCF;][]{hahn2020bayesian} model to analyze multivariate response data for causal inference,
and \citet{esser2024seemingly} proposed a method of incorporating seemingly unrelated regression \citep[SUR;][]{zellner1962efficient} into the multivariate BART framework.

Given a fully observed set of responses $\mathbf{Y}_i$ and model covariates $\mathbf{X}_i$, the multivariate BART model can be formulated as
\begin{equation} \label{eq: mvBART}
    \mathbf{Y}_i = \sum_{k=1}^K g\left(\mathbf{X}_i; \mathcal{T}_k, \mathbf{Q}_k\right) + \boldsymbol{\epsilon}_i, \quad \boldsymbol{\epsilon}_i \overset{i.i.d.}{\sim} \mathcal{N}_p \left(\mathbf{0}, \boldsymbol{\Omega}^{-1} \right),
\end{equation}
where $\boldsymbol{\Omega}$ is the $p \times p$ residual precision matrix, $K$ is the total number of regression trees employed in the model, $\mathcal{T}_k$ is the $k^\text{th}$ regression tree, and $\mathbf{Q}_k$ contains the $p$-dimensional node-specific vectors $\boldsymbol{\mu}_{k\ell} = (\mu_{1k\ell},\ldots,\mu_{pk\ell})$. 

We calibrate the priors of the multivariate BART model by extending the calibration techniques adopted by \citet{chipman2010bart} and using the same prior specifications. 
%The tree priors $p(\mathcal{T}_k)$ for all $k$ trees remain unchanged.
The prior for $\boldsymbol{\mu}_{k\ell}$ is assigned a $p$-variate normal distribution, $\boldsymbol{\mu}_{k\ell} \sim \mathcal{N}_p(\boldsymbol{\mu}_{\mu}, \boldsymbol{\Omega}_{\mu}^{-1})$ with $\boldsymbol{\Omega}_{\mu} = \tau_\mu \mathbf{I}_p$, where $\mathbf{I}_p$ is the $p$-dimensional identity matrix.
In the univariate BART framework from \citet{chipman2010bart}, the response is scaled and shifted to the range $\lbrack -0.5, 0.5\rbrack$, justifying the setting of $\mu_{\mu} = 0$. 
The prior precision $\tau_\mu$ is calibrated based on some prior probability $\rho_\mu$ that $\mathbb{E}(Y_i \given\mathbf{X_i})$ falls inside this rescaled interval.
We calibrate $\boldsymbol{\mu}_{\mu}$ and $\tau_\mu$ as per the univariate setting.
Although the prior on $\mathbf{\Omega}_\mu$ assumes no covariance between the components of $\boldsymbol{\mu}_{k\ell}$, the posterior distribution of $\mathbf{\Omega}_\mu$ is expected to be non-diagonal when there is information to be shared across response variables.

For $\boldsymbol{\Omega}$, we assign a conjugate Wishart prior such that $\boldsymbol{\Omega} \sim \mathcal{W}_p \left(\nu, \mathbf{V}\right)$. 
In the univariate setting, the residual precision $\tau$ is assigned a $\operatorname{Ga}(\frac{\nu}{2}, \frac{\nu \lambda}{2})$ prior. 
A value for $\nu$ is first chosen based on the shape of the prior curve.  
Next, a rough data-based under-estimate $\hat{\tau}$ is obtained, either via the sample precision or the estimated residual precision from a least squares linear regression model, with the latter being our default.
With the assumption that the BART model estimates a residual precision $\tau$ at least as large as the rough estimate $\hat{\tau}$ with prior probability $\rho_\tau$, the hyperparameter $\lambda$ can be obtained from $P\left(\tau > \hat{\tau}\right) = \rho_\tau$ after selecting appropriate values for $\nu$ and $\rho_\tau$.
The default setting for $(\nu,\rho_\tau)$ from \citet{chipman2010bart} is $(3, 0.9)$.
Given that the Wishart distribution is a multivariate extension of the gamma distribution, we calibrate the hyperparameters $\nu$ and $\mathbf{V}$ by first choosing a value for $\nu$ and a vector of $p$ probabilities $\boldsymbol{\rho}_\tau = (\rho_{\tau 1},\ldots,\rho_{\tau p})$ 
as in the univariate BART case. 
The rough data-based estimates $\hat{\tau}_1,\ldots,\hat{\tau}_p$ are then obtained for each univariate column of the outcome $\mathbf{Y}$. The scale matrix of the Wishart prior is then given by $\mathbf{V} = (\nu \boldsymbol{\lambda})^{-1} \mathbf{I}_p$, where $\boldsymbol{\lambda} = (\lambda_1,\ldots,\lambda_p)$, and each $\lambda_j$ is calculated as in the univariate case. 

Extending the probit BART model to the multivariate framework follows a similar data-augmentation approach as in the univariate setting; given binary outcomes $\mathbf{Y}_i$, the latent variables $\mathbf{Y}^\star_i$ follow a multivariate BART model such that 
\begin{equation} \label{eq: mvBART probit}
    \mathbf{Y}^\star_i = \sum_{k=1}^K g\left(\mathbf{X}_i; T_k, \mathbf{Q}_k\right) + \boldsymbol{\epsilon}_i, \quad \boldsymbol{\epsilon_i} \overset{i.i.d.}{\sim} \mathcal{N}_p\left(\mathbf{0},\mathbf{R}\right).
\end{equation}
where $\mathbf{R}$ is constrained to be a correlation matrix, as before.

\section{Joint Models for Multivariate MNAR Missing Data} 
\label{sec: 5 Joint models}
We now describe our two joint models, missBART1 and missBART2, developed under the selection model framework from Equation \eqref{eq: selection model factorization} to handle data with missing responses.
While both models assign the multivariate BART function from Equation \eqref{eq: mvBART} to the data model $\Tilde{\mathbf{Y}}\given \mathbf{X}, \boldsymbol{\theta}$, missBART1 assigns a multivariate probit regression model from Section \ref{sec: 3 Bayesian probit regression} to the missingness model $\mathbf{M}\given \mathbf{X}, \Tilde{\mathbf{Y}}, \boldsymbol{\psi}$. By contrast missBART2 assigns the probit BART model as the missingness model instead.
Of key importance in these models is that $\mathbf{X}$ is used as the covariates in the BART data model, and the set $(\mathbf{X}, \Tilde{\mathbf{Y}})$ is used as the predictors in the missingness model. 
Details of both models are given in the framework of multivariate responses below. It is straightforward to reduce both models to their univariate equivalents.

\subsection{missBART1} \label{subsec: missBART1}
The complete data likelihood under this model is:
\begin{equation}
    p\left(\Tilde{\mathbf{Y}}, \mathbf{M}\given\mathbf{X}, \bm{\mathcal{T}}, \mathbf{Q}, \boldsymbol{\Omega}, \mathbf{B}, \mathbf{R}\right) = \underbrace{p\left(\Tilde{\mathbf{Y}}\given\mathbf{X}, \bm{\mathcal{T}}, \mathbf{Q}, \boldsymbol{\Omega}\right)}_{\text{BART regression model}} %\times 
    \underbrace{p\left(\mathbf{M}\given\mathbf{Z}, \mathbf{B}, \mathbf{R}\right)}_{\substack{\text{probit regression} \\ \text{missingness model}}},
    \label{eq:missBART1}
\end{equation}
where $\mathbf{Z} = (\mathbf{1}, \mathbf{X}, \Tilde{\mathbf{Y}})$.
This specification accounts for MCAR, MAR, and MNAR mechanisms based on the dependence of detection probabilities on covariates or the response itself.
The matrix of coefficients $\mathbf{B}$ has dimensions $r \times p$ where $r=1+p+q$ is the total number of columns in $\mathbf{Z}$, including the intercept. 
Defining $\mathbf{B}^{(j)} = (\mathbf{B}_{1j},\ldots, \mathbf{B}_{rj})$ as the $j^{\text{th}}$ column of $\mathbf{B}$, each element of $\mathbf{B}^{(j)}$ represents the degree to which each predictor influences the missingness of response $j$.

While the multivariate probit regression model from \citet{talhouk2012efficient} assumes that $\boldsymbol{\Psi}$ in Equation \eqref{eq: prior B} is known, we propose a modification which allows for the incorporation of prior beliefs about the missing data mechanism. We note that small values on the diagonal of $\boldsymbol{\Psi}$ correspond to probit regression coefficients that are close to zero, and thus have little effect on the missingness. Since our missingness model predictors consist of $\mathbf{X}$ and $\Tilde{\mathbf{Y}}$, we can tailor the prior distribution to favor MAR (coefficients associated with $\Tilde{\mathbf{Y}}$ close to zero) or MNAR structures (coefficients associated with $\Tilde{\mathbf{Y}}$ are not zero). 

To structure this prior distribution, first we set $\boldsymbol{\Psi}^{-1} = \operatorname{diag}(\tau_{B_0}, \tau_{B_{X}} \boldsymbol{1}_q, \tau_{B_{Y}}\boldsymbol{1}_p)$, where $\boldsymbol{1}_q$ and $\boldsymbol{1}_p$ are $q$ and $p$-dimensional vectors of ones, respectively.
Next, we assign separate gamma priors to $\tau_{B_0}, \tau_{B_{X}}$, and $\tau_{B_{Y}}$ with shape and rate parameters $(\alpha_0, \beta_0)$, $(\alpha_X, \beta_X)$, and $(\alpha_Y, \beta_Y)$.
See \ref{Appendix: posterior dist of Psi} for the full derivation of the posterior distributions.
By carefully tuning these prior parameters, we can articulate any prior beliefs regarding the missingness mechanisms. 
As a default, we set $(\alpha_0, \beta_0) = (2,1)$, $(\alpha_X, \beta_X) = (1+q, 1)$ and $(\alpha_Y, \beta_Y) = (1+p+q, 1)$.
Scaling the prior mean of $\tau_{B_{Y}}$ with the number of covariates and responses ensures that it remains larger than the prior mean of $\tau_{B_{X}}$, increasing the likelihood that the model is \emph{a priori} MAR. 
This also helps keep the values in $\mathbf{B}$ small as the number of covariates and responses increases, and \emph{vice versa}.

Under the data augmentation scheme from Equation \eqref{eq: mv probit latent}, the joint posterior distribution of the model takes the form
\begin{equation}
    p\left(\bm{\mathcal{T}}, \mathbf{Q}, \boldsymbol{\Omega}, \mathbf{M}^\star, \mathbf{Y}^{mis}, \mathbf{B}, \mathbf{R}, 
\boldsymbol{\Psi} \given \mathbf{X}, \mathbf{Y}^{obs}, \mathbf{M}\right).
\end{equation}
By assigning the priors specified previously, posterior sampling for $\bm{\mathcal{T}}$ is carried out via Metropolis-Hastings, while all other parameters have Gibbs updates.
This allows for posterior inferences to be made on $\mathbf{Y}^{mis}$ and $\mathbf{B}$. 
Thus, imputations for missing responses can be obtained via computing the posterior mean of $\mathbf{Y}^{mis}$, along with uncertainty quantification. 
Further, posterior intervals of $\mathbf{B}^{(j)}$ --- to the extent that they include or exclude $0$ --- can give insights into the underlying missing data mechanism of the partially-observed data. 
The posterior sampling algorithm for missBART1 is outlined in \ref{Appendix: sampling missBART1}.

\subsection{missBART2} \label{subsec: missBART2 model}
Due to its inherent linear structure, the probit regression model may prove inadequate when the true underlying relationship in the missing data model is non-linear.
Additionally, the missing data model specification in missBART1 excludes any interaction terms between the model predictors. 
Consequently, to account for potential non-linearity and predictor interactions, we propose a more flexible, fully non-parametric selection model by applying a probit BART model to the missingness mechanism model. Thus, the model effectively amounts to replacing the probit regression missingness model in Equation \eqref{eq:missBART1} with a multivariate probit BART model. 
The complete data likelihood for missBART2 is
\begin{equation}
    p\left(\Tilde{\mathbf{Y}}, \mathbf{M}\given\mathbf{X}, \bm{\mathcal{T}}^{y}, \mathbf{Q}^{y}, \bm{\mathcal{T}}^{m}, \mathbf{Q}^{m}, \boldsymbol{\Omega}, \mathbf{R}\right) = \underbrace{p\left(\Tilde{\mathbf{Y}}\given\mathbf{X}, \bm{\mathcal{T}}^{y}, \mathbf{Q}^{y}, \boldsymbol{\Omega}\right)}_{\text{BART regression model}} %\times 
    \underbrace{p\left(\mathbf{M}\given\mathbf{Z}, \bm{\mathcal{T}}^{m}, \mathbf{Q}^{m}, \mathbf{R}\right)}_{\substack{\text{probit BART} \\ \text{missingness model}}},
\end{equation}
where here the intercept is omitted from $\mathbf{Z}$.
Overall, the model uses two distinct sets of trees with associated terminal node parameters, denoted by $(\bm{\mathcal{T}}^{y}, \mathbf{Q}^{y}) = (\mathcal{T}^{y}_1, \mathbf{Q}^{y}_1), \ldots, (\mathcal{T}^{y}_{K_y}, \mathbf{Q}^{y}_{K_y})$ for the regression on $\Tilde{\mathbf{Y}}$ and similarly
for $\mathbf{M}$, where $K_y$ and $K_m$ are the respective numbers of trees in each BART model.

In \citet{esser2024seemingly}, a parameter-expanded data augmentation technique from \cite{zhang2020parameter} is adopted for estimating $\mathbf{R}$ from Equation \eqref{eq: mvBART probit}.
Given the flexibility of tree structures in the missing data model to capture the underlying signal, we find it sufficient to fix $\mathbf{R} = \mathbf{I}_p$.
This choice also reduces the computational burden of estimating the correlation structure via additional data augmentation techniques.
Furthermore, our simulation studies show that even when the missing data model is simulated using a non-diagonal correlation matrix, missBART2 still performs well despite this simplification.

While missBART1 has the flexibility of incorporating prior knowledge of the missing data mechanisms via the prior calibration of $\mathbf{B}$, the probit BART model in missBART2 includes both $\mathbf{X}$ and $\Tilde{\mathbf{Y}}$ as predictors available to be used in the splitting rules of $\bm{\mathcal{T}}^m$, leveraging the automatic variable selection feature of BART to recover the missing data mechanism without needing to pre-specify the functional form of the probit model.
Although missBART2 does not have the luxury of being able to examine interpretable probit regression parameters, identifying which variables in $\mathbf{X}$ and $\Tilde{\mathbf{Y}}$ are most commonly used to construct splitting rules in the missingness model's set of trees can nonetheless provide an indication of whether the missingness mechanism depends only on observed quantities (i.e., MAR) or unobserved quantities (i.e., MNAR).
Simulation studies from \citet{chipman2010bart} on BART variable selection show that employing fewer trees can lead to better identification of important and influential variables.
As the number of trees increases, the frequency with which each variable is used in the splitting rules becomes more uniform. 
However, too few trees can hinder convergence and compromise overall model fit.
Thus, care should be taken when defining the number of trees to be used in the missingness model, particularly when the recovery of different missing data mechanisms is of interest.

An additional advantage can be gained by incorporating the \texttt{BARTm} methodology introduced in \citet{kapelner2015prediction} to missBART2 such that it can effectively handle missing values in $\mathbf{X}$. 
This approach seamlessly integrates covariate missingness into the tree-splitting rules, enabling splits based on the available $\mathbf{X}$ values as well as their missingness status. 
In contrast, the parametric constraints of the probit regression model necessitate prior covariate imputation in missBART1 when faced with missing values in $\mathbf{X}$.

Following the latent variable transformation of $\mathbf{M}$ to $\mathbf{M}^\star$, the joint posterior distribution takes the form
\begin{equation}
    p\left(\bm{\mathcal{T}}^{y}, \mathbf{Q}^{y}, \bm{\mathcal{T}}^{m}, \mathbf{Q}^{m}, \boldsymbol{\Omega}, \mathbf{M}^\star, \mathbf{Y}^{mis}\given \mathbf{X}, \mathbf{Y}^{obs}, \mathbf{M}\right).
\end{equation}
The sampling procedure for missBART2 is similar to that for missBART1, but with a few exceptions. 
First, posterior sampling for both sets of trees $\bm{\mathcal{T}}^{y}$ and $\bm{\mathcal{T}}^{m}$ are carried out via Metropolis-Hastings steps.
Next, $\mathbf{Q}^{y}, \boldsymbol{\Omega}, \mathbf{M}^\star$, and $\mathbf{Q}^{m}$ are updated via Gibbs.
No sampling is required for $\mathbf{B}$ and $\mathbf{R}$ since the BART model is non-parametric, and we fix $\mathbf{R}=\mathbf{I}_p$ so the parameter expansion technique from \citet{talhouk2012efficient} is not required. 

Defining the regression model parameters $\boldsymbol{\theta}=\{\bm{\mathcal{T}}^y,\mathbf{Q}^y,\boldsymbol{\Omega}\}$ and missing model parameters $\boldsymbol{\psi}=\{\bm{\mathcal{T}}^m,\mathbf{Q}^m\}$, the full conditional distribution of $\mathbf{Y}^{mis}$ is
$
        p({\mathbf{Y}}^{mis} \given 
        {\mathbf{Y}}^{obs}, \mathbf{M}^\star, \boldsymbol{\theta}, \boldsymbol{\psi}) \propto p({\mathbf{Y}^{obs}, \mathbf{Y}}^{mis} \given \boldsymbol{\theta}) p(\mathbf{M}^\star \given \mathbf{Y}^{obs}, \mathbf{Y}^{mis},\boldsymbol{\psi})$,
where no known distributional form is available, necessitating the implementation of a Metropolis-Hastings step. 
For each missing entry in iteration $t$, we propose a new value $Y^{mis}_{t,i,j}$ from a random-walk proposal distribution $\mathcal{N}(Y^{mis}_{t-1,i,j}, \sigma_Y^2)$. Given that the observed data are scaled to the range $\lbrack-0.5, 0.5\rbrack$, we set $\sigma_Y=0.5/p$ to ensure that proposed values of $\mathbf{Y}^{mis}$ do not go too far outside this range. This tuning choice has demonstrated good empirical performance in simulations.
The posterior sampling algorithm for missBART2 is outlined in \ref{Appendix: sampling missBART2}.

\section{Simulation Study}
\label{sec: 6 Simulation Study}
We present the results from our simulation studies aimed at demonstrating the advantages of missBART1 and missBART2 for performing predictive tasks in the presence of non-ignorable missingness. 
We highlight two scenarios: bivariate MAR and MNAR data, and multivariate MNAR data with missing covariates. Two univariate examples are included in \ref{Appendix: univariate examples}.
For each scenario, we compare missBART1 and missBART2 with $6$ other methods: multivariate BART on complete cases (`mvBART\_cc'), multivariate BART on the \texttt{missForest} imputed data (`mvBART\_imp'), univariate BART on the complete cases of each response variable (`uniBART\_cc'), univariate BART on each \texttt{missForest} imputed response (`uniBART\_imp'), as well as missBART1 and missBART2 on each univariate response variable (`uni\_missBART1', `uni\_missBART2').
We run each model for $5000$ burn-in and $5000$ post-burn-in iterations. 
The number of trees in the data model is fixed at $100$.
In the bivariate examples, $20$ probit BART trees are used in missBART2.
In the multivariate examples, however, more response variables and covariates are present in the missingness model, as well as induced missingness in the covariates. 
Given the more challenging simulation setup, we increase the number of missingness trees to $50$ to allow for better model convergence without sacrificing the variable selection capabilities of the probit BART trees.

\subsection{Bivariate Examples: Missingness Under MAR and MNAR}
\label{subsec: bivariate examples}
We now demonstrate the performance of missBART1 and missBART2 on simulated bivariate data under two MAR and two MNAR scenarios. 
We simulate $n=2000$ i.i.d bivariate observations from a multivariate BART model with $q=5$ covariates. 
Using the same complete data, we create four different scenarios by varying the missing data mechanism (MAR or MNAR) and the underlying missing data model --- multivariate probit regression (MAR 1 and 2) or multivariate probit BART (MNAR 1 and 2).
In both MAR settings, only $\mathbf{X}$ is used to simulate missingness. In contrast the missingness probabilities only depend on $\mathbf{Y}$ for both MNAR settings.
A summary of the simulation details is given in \ref{Appendix: bivariate simulation details}. 

\begin{figure}[H]
    \centering
    \includegraphics[width=\linewidth, trim={0 0.55cm 0 0}, clip]{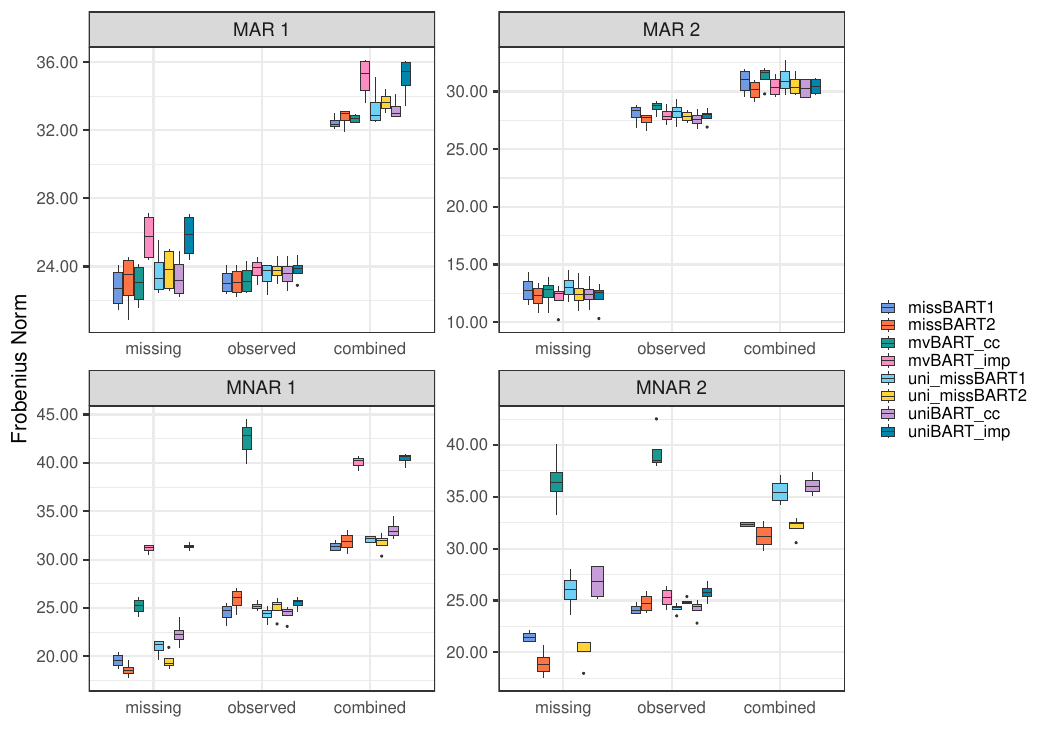}
    \caption{Frobenius norms from the bivariate simulated MAR and MNAR scenarios. The Frobenius norms are calculated using the true values of the missing, observed, and combined responses.}
    \label{fig: bivariate Frobnorms}
\end{figure}

Through $4$-fold cross-validation, we make comparisons between the eight models.
For mvBART\_imp and uniBART\_imp, missing $\mathbf{Y}$ imputations are first obtained by passing the entire $(\mathbf{X}, \mathbf{Y})$ to the \texttt{missForest} function, using its default settings, before splitting $\mathbf{Y}$ into train/test sets. 
As true values of the simulated $\mathbf{Y}$ prior to imposing missingness are available, we can evaluate both imputation and prediction accuracy by separately computing Frobenius norms for the missing, observed, and combined responses. The out-of-sample Frobenius norms for all four examples are shown in Figure \ref{fig: bivariate Frobnorms}. Some results were omitted due to overly high values, namely mvBART\_cc for the combined data in MNAR 1, mvBART\_imp and uniBART\_imp for the missing data in MNAR 2, and mvBART\_cc and uniBART\_imp for the combined data in MNAR 2.

As expected, the joint models perform no worse than the complete-case and \texttt{missForest}-imputed models for both missing and observed responses in MAR settings, as the missing data mechanism can be ignored and only the regression model is required. The MNAR cases show wider disparities between the models, especially for missing responses. This was also expected, as complete-case and \texttt{missForest}-imputed models fail to account for any relationships between responses and their corresponding missingness status. 
missBART1 only outperforms missBART2 in the MAR 1 case, where missingness was generated from probit regression rather than trees. The poor performance of \texttt{missForest} imputation in all but the MAR 2 case is particularly notable. This may be due to inaccurate imputations obtained prior to model fitting, leading to erroneous results. In addition, the outcome-specific out-of-sample root mean squared errors and continuous ranked probability scores 
\citep{gneiting2007strictly}, which we defer to \ref{Appendix: bivariate RMSE and CRPS}, also align with these findings.

\begin{figure}[H]
    \centering
    \begin{subfigure}[b]{0.49\textwidth}
        \centering
        \includegraphics[width=\linewidth, trim={0 0.35cm 0 0}, clip]{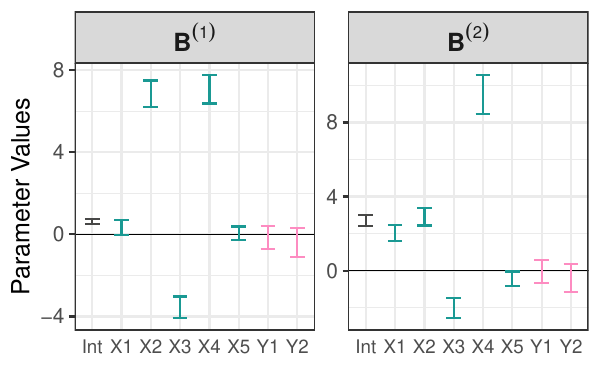}
        \caption{MAR 1: The posterior intervals of $\mathbf{B}_Y$ overlap with 0, capturing the true MAR mechanism.}
        \label{fig: MAR 1 B_post}
    \end{subfigure}
    \hfill 
    \begin{subfigure}[b]{0.49\textwidth}
        \centering
        \includegraphics[width=\linewidth, trim={0 0.35cm 0 0}, clip]{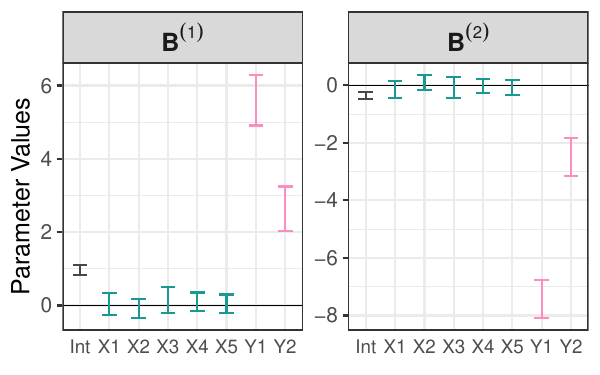}
        \caption{MNAR 1: The posterior intervals of $\mathbf{B}_Y$ do not contain $0$, capturing the true MNAR mechanism.}
        \label{fig: MNAR 1 B_post}
    \end{subfigure}
    \caption{$95\%$ posterior intervals of $\mathbf{B}$ from missBART1 for MAR 1 and MNAR 1. 
    \label{fig: B posterior interval}}    
\end{figure}

To assess recovery of the underlying missing mechanism, we examine the posterior intervals of $\mathbf{B}$ from missBART1.
Figure \ref{fig: B posterior interval} shows the $95\%$ posterior intervals for the coefficients of the intercept, five covariates, and two responses, within the probit regression model of missBART1 for MAR 1 and MNAR 1. 
Notably, the error bars representing coefficients in $\mathbf{B}_Y$ overlap with 0 when the data are MAR, correctly indicating the absence of a relationship between missingness and the responses.
In contrast, the intervals do not overlap with 0 in the MNAR scenario, successfully capturing the non-ignorable missing data mechanism.

\begin{figure}[H]
    \centering
    \includegraphics[width=\linewidth, trim={0 0.35cm 0 0}, clip]{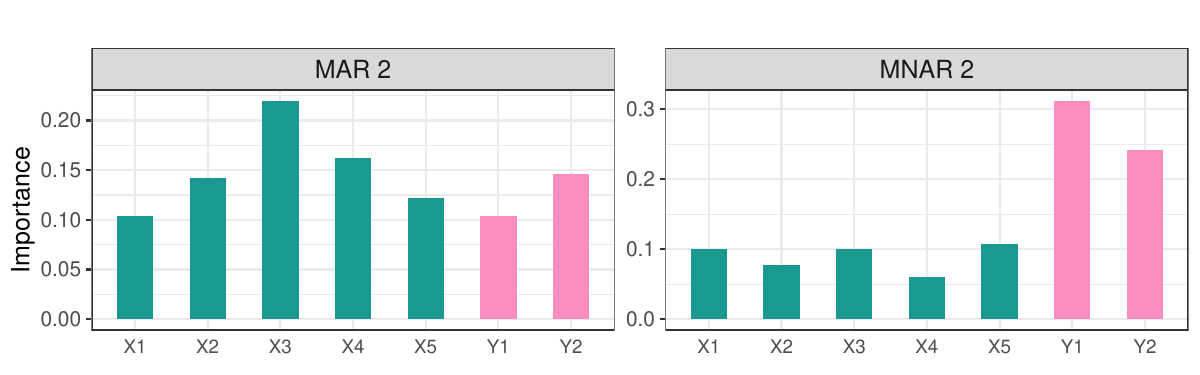}
    \caption{Variable importance from the missingness trees of missBART2 for MAR 2 and MNAR 2. Covariates are most frequently used in the splitting rules under MAR 2, while there is a clear distinction between the importance of responses and covariates under MNAR 2.}
    \label{fig: MAR 2/MNAR 2 varimp}
\end{figure}

To assess the recovery of the missing data mechanism from missBART2, we investigate the variable importance\footnote{Calculated as the average number of uses of each variable in the splitting rules of the missingness trees over $5000$ post-burn-in iterations.} of its missingness trees under MAR 2 and MNAR 2
in Figure \ref{fig: MAR 2/MNAR 2 varimp}. 
For MNAR 2, most splits are made on $\mathbf{Y}^{(1)}$ and $\mathbf{Y}^{(2)}$, while only a small proportion of splits are made on the covariates. 
In MAR 2, $\mathbf{X}^{(3)}$ is most frequently used, followed by $\mathbf{X}^{(4)}$.

\subsection{Multivariate Examples: MNAR Response, MAR Covariates}
\label{subsec: multivariate examples}
While there are no missing covariates in the \textit{global Amax} data, we present three simulations (MNAR\_amp0, MNAR\_amp1, and MNAR\_amp2) with MNAR missingness in multivariate responses and different missingness in the covariates to show the performance of our joint models in the presence of ignorably missing covariates and non-ignorably missing outcomes. We use the same data, keep the missingness in $\mathbf{Y}$ consistent throughout, but vary the missingness in $\mathbf{X}$. We first create data with $n=2000$ i.i.d observations with $p=5$ responses from the multivariate version of the Friedman function \citep{friedman1991multivariate}:
\begin{equation}
    \label{eq: mv Friedman function}
    \mathbf{Y}_i =  \boldsymbol{\xi}_{1} \sin{(\pi X_{i1} X_{i2})} + \boldsymbol{\xi}_{2} (X_{i3} - 0.5)^2 + \boldsymbol{\xi}_{3} X_{i4} + \boldsymbol{\xi}_{4} X_{i5} + \boldsymbol{\epsilon}_i, \quad \boldsymbol{\epsilon}_i \overset{i.i.d}{\sim} N_p(\mathbf{0}, \boldsymbol{\Sigma}_{\epsilon})
\end{equation}
where $\boldsymbol{\xi}_1,\ldots, \boldsymbol{\xi}_4$ are $p$-dimensional vectors of coefficients drawn from $\mathcal{N}_p(\mathbf{0}, \boldsymbol{\Sigma}_\xi)$.
While the Friedman function only requires five covariates, 
we include five additional non-informative ones. Each covariate is simulated independently from a continuous $\text{Unif}(0,1)$ distribution.

Previously, missingness either depended strictly on $\mathbf{X}$ in MAR cases or $\mathbf{Y}$ for MNAR. Here, the missingness model is a multivariate probit regression with non-zero coefficients for $(\mathbf{X}, \mathbf{Y})$, i.e. missingness in $\mathbf{Y}$ depends on both observed and unobserved variables. 
However, to ensure the data are MNAR, we enforce stronger relationships between the missingness probabilities and $\mathbf{Y}$ while keeping the coefficients of $\mathbf{X}$ near $0$.

As a baseline, the complete set of covariates is used in the first example, MNAR\_amp0, such that missingness is only present in the responses. 
For MNAR\_amp1 and MNAR\_amp2, MAR missingness is introduced into $\mathbf{X}$ using the function \texttt{ampute} \citep{schouten2018generating} from the \texttt{mice} package, which allows specifying the underlying missing data mechanism, the overall proportion of missingness in $\mathbf{X}$, and the missing data pattern \citep{van2018flexible}. 
For both examples, we use the default setting of \texttt{"MAR"} for the missing data mechanism in the covariates and $0.5$ for the associated proportion of missingness.
By default, the missing data pattern is ``diagonal'', such that only one variable is missing in each row $\mathbf{X}_i$.
We use this setting in MNAR\_amp1, keeping the missingness proportion across each covariate relatively consistent (between $4.35\%$ and $5.65\%$ missing). 
In MNAR\_amp2, we allow for more complicated patterns of covariate missingness with varying proportions of missingness (between $5.70\%$ and $39.65\%$).
The missingness patterns for MNAR\_amp1 and MNAR\_amp2 are included in \ref{Appendix: missing X patterns}. 

We use $8$-fold cross-validation to compare the models described in Section \ref{subsec: bivariate examples} and an additional model missBART2\_impX, the missBART2 model with prior covariate imputation. 
For MNAR\_amp0, since the covariates are fully observed and no covariate imputation is necessary, missBART2\_impX is equivalent to missBART2 and thus excluded from the study.
Note that, while missBART1, missBART2\_impX, and uni\_missBART1 are trained and tested on the \texttt{missForest} imputed covariates, all other models include the \texttt{BARTm} method for handling missing covariates within the BART trees. 
The out-of-sample Frobenius norms of the three examples are shown in Figure \ref{fig: MNAR_amp}, while the out-of-sample RMSE plots are included in \ref{Appendix: MNAR_amp RMSE}.
\begin{figure}[H]
    \centering
    \includegraphics[width=\linewidth, trim={0 0.7cm 0 0}, clip]{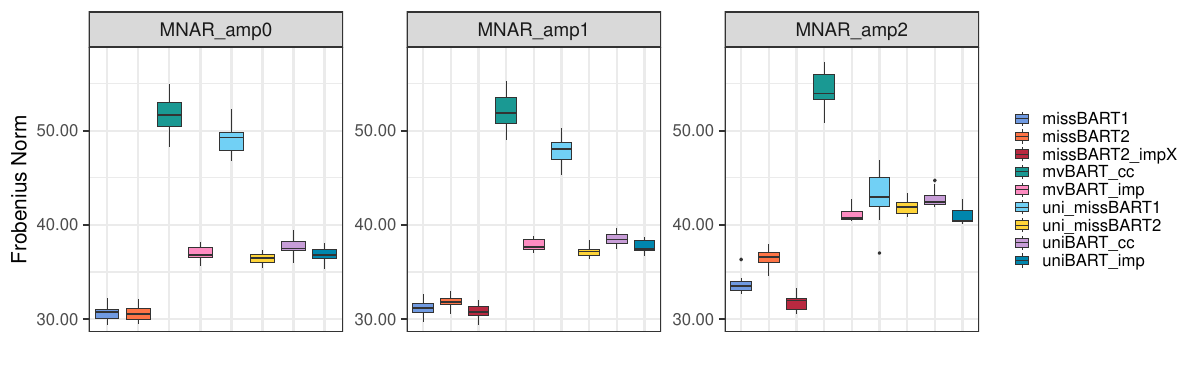}
    \caption{Out-of-sample Frobenius norms for the three MNAR examples with missing covariates.}
    \label{fig: MNAR_amp}
\end{figure}%
The joint models again demonstrate robust performance when dealing with multivariate non-ignorable responses, both in the presence and absence of missing covariates. 
When covariates are fully observed, missBART1 and missBART2 yield comparable results, outperforming all other methods.
However, the introduction of missingness in the covariates generally leads to a decline in model performance. In the MNAR\_amp1 scenario with ``diagonal'' missing covariates, both joint models show increased Frobenius norms compared to the fully observed case, with missBART1 slightly outperforming missBART2. missBART2\_impX, which is essentially a two-step process that requires covariate imputation to be applied prior to model fitting, emerges with the strongest overall performance. However, missBART2 with the incorporation of the \texttt{BARTm} technique is a more straightforward approach to handling missing covariates and its performance is competitive with missBART2\_impX. In the more complex MNAR\_amp2 scenario, the disparity in performance between missBART1, missBART2, and missBART2\_impX becomes more distinct, yet missBART2\_impX maintains the best overall performance, showing a lower Frobenius norm than missBART1, in particular, which is also based on prior covariate imputation.

\section{Application: \textit{global Amax}}
\label{sec: 7 Real data example}
The \textit{global Amax} data, originating from \citet{maire2015global}, comprises a rich set of environmental covariates, including 20 soil and 26 climate variables, alongside 5 continuous responses reflecting leaf photosynthetic traits: light-saturated photosynthetic rate (\textit{Aarea}), stomatal conductance (\textit{Gs}), leaf nitrogen (\textit{Narea}), leaf phosphorus (\textit{Parea}), and specific leaf area (\textit{SLA}). 
For details, see Appendix S2 and S3 of \citet{maire2015global}. 
The authors used various univariate linear and parametric methods such as mixed-effects regression models, variation partitioning, redundancy analysis, and path analysis to quantify the influence of climate and soil properties on individual leaf photosynthetic traits.
The study identified key environmental variables, particularly soil pH (\textit{pH}), moisture index (\textit{Miq}), and available soil phosphate content (\textit{Pavail}), as major influences on photosynthetic traits. 

However, the methods employed by \citet{maire2015global} lack the flexibility of non-parametric BART models in capturing non-linear relationships, and do not address the partially overlapping nature of missingness in the multivariate response.
In fact, there was minimal discussion of how missingness was addressed, suggesting an implicit reliance on the ignorability assumption.
For further details on the methods carried out in this previous study, see Appendix S6 of \citet{maire2015global}.

\subsection{Missingness in \textit{global Amax} data}
While all 46 covariates in the \textit{global Amax} data are fully observed, the responses exhibit a high level of missingness with various missingness patterns. 
Specifically, out of 2368 rows, only 217 ($\approx 9.16\%$) are complete. 
An illustration for the different missing data patterns identified in the responses are shown in \ref{Appendix: missing patterns response}.
Among the 5 responses, \textit{Aarea} is mostly complete, with only 12 ($\approx 0.5\%$) missing cases, while \textit{SLA} and \textit{Narea} have moderate missingness at 433 ($\approx 18.29\%$) and 652 ($\approx 27.53\%$) cases, respectively.
In contrast, \textit{Gs} and \textit{Parea} exhibit the highest levels of missingness, with 1353 ($\approx 57.14\%$) and 1836 ($\approx 77.5\%$) missing cases, respectively. These patterns likely stem from the compilation of the \textit{global Amax} data by \citet{maire2015global}, which focused primarily on \textit{Aarea}, a key photosynthetic trait, explaining its low level of missingness. 
Traits such as \textit{SLA} and \textit{Narea}, which are frequently quantified in studies assessing \textit{Aarea}, also show relatively low missingness.
In contrast, the higher levels of missingness observed for \textit{Parea} and \textit{Gs} may reflect their more complex or less frequent measurement in the contributing studies. Despite the substantial amount of missing responses in the data, \citet{maire2015global} offered limited details regarding how missingness was managed in their analyses. The redundancy analysis briefly noted the exclusion of \textit{Parea} due to its small sample size, resulting in a dataset of 647 species, suggesting an approach that relied on complete cases.

\subsection{Results}
\begin{figure}[H]
    \centering
    \includegraphics[height=\dimexpr\textheight - 4\baselineskip\relax]{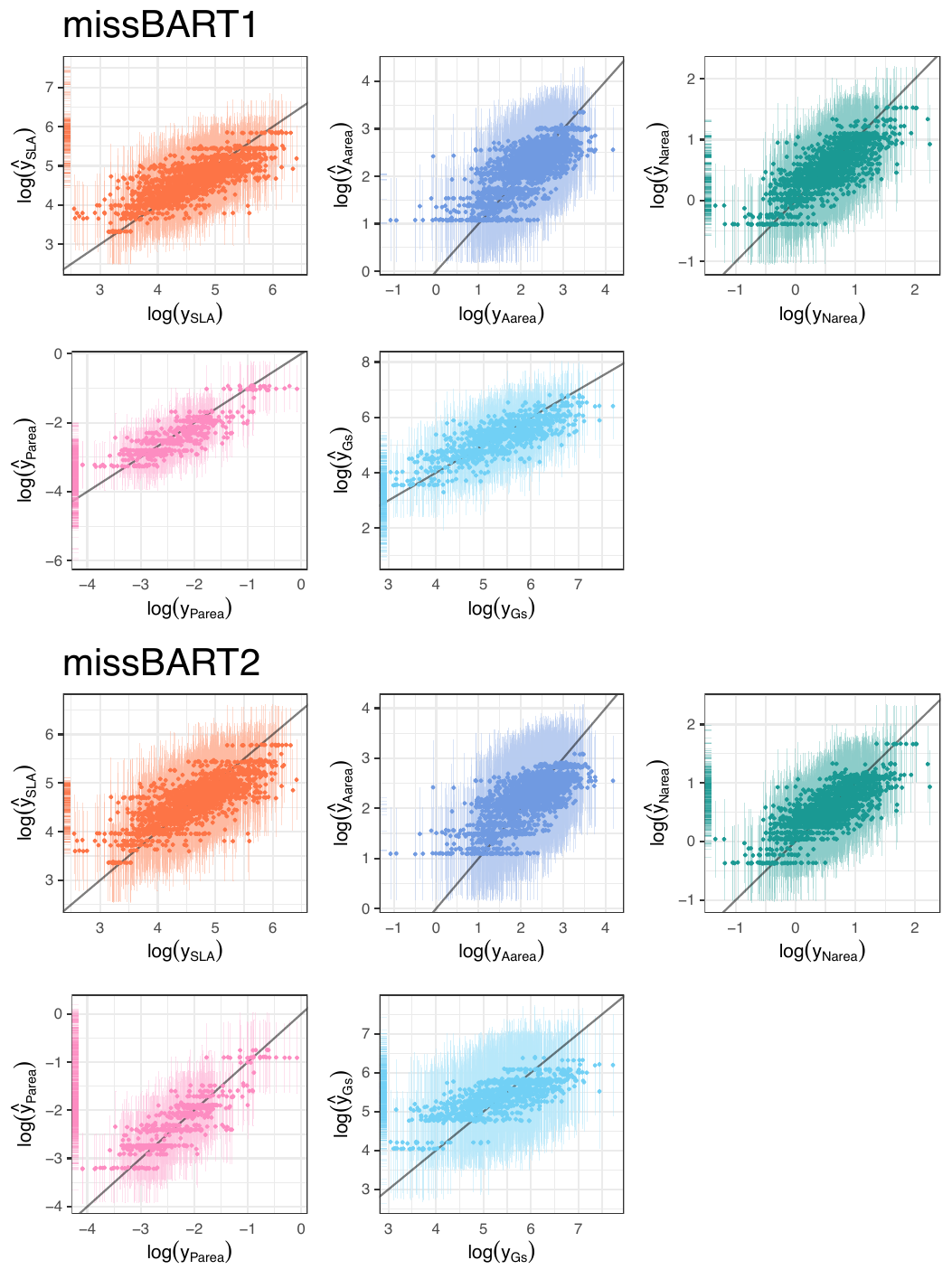}
    \caption{missBART1 and missBART2 predictions for the observed data against their true log-transformed values. Vertical error bars represent the $95\%$ prediction intervals for the observed data. Rug plots on the $y-$axes show the posterior means of the missing data imputations.} 
    \label{fig: ypred missBART1and2}
\end{figure}

We now apply missBART1 and missBART2 to the \textit{global Amax} data. 
Prior to model fitting, a log transformation is applied to each response variable to account for right-skewness, as per \citet{maire2015global}.
We use $5000$ burn-in and $5000$ post-burn-in iterations, $100$ regression trees, and $50$ missingness trees. Figure \ref{fig: ypred missBART1and2} shows the model predictions for the observed data against their true log-transformed values for missBART1 and missBART2, respectively, along with vertical error bars representing their $95\%$ prediction intervals. 
Rug plots on the $y$-axis display the posterior means of the missing value imputations, providing insight into both models' imputations for the missing data. 
Both models demonstrate strong predictive performances on the observed data, indicating their robustness in making accurate predictions for observed cases, while simultaneously imputing the missing values.
Notably, despite similar fits on the observed data, the imputations differ between the models. 
While most imputations from missBART2 closely align with the range of observed values, missBART1 shows more drastic deviations, especially for \textit{SLA}, \textit{Parea}, and \textit{Gs}.

\begin{figure}[H]
    \centering
    \includegraphics[width=\linewidth]{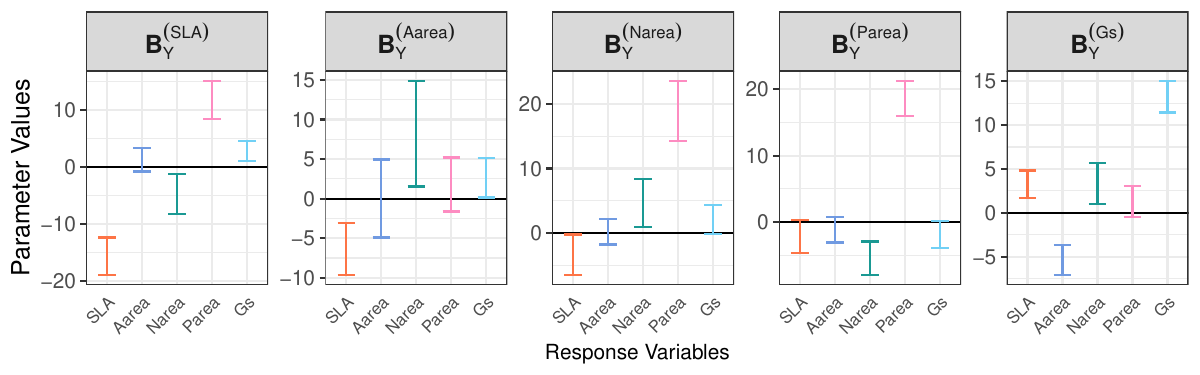}
    \caption{$95\%$ posterior intervals of $\mathbf{B}_Y$, the probit regression coefficients associated with the response variables from missBART1. Most intervals do not overlap with 0, indicating a strong presence of MNAR missingness.}
    \label{fig: plant B_post}
\end{figure}

Figure \ref{fig: plant B_post} shows the 95\% posterior intervals of $\mathbf{B}_Y$ obtained from missBART1.
In each panel, the error bars indicate the degree to which the missingness of each response variable is influenced by the values of the 5 response variables.
The coefficients for \textit{Narea} remain consistently non-zero (do not overlap with zero) across multiple panels, apart from that for \text{Gs}, indicating that \textit{Narea} strongly influences the missingness probabilities for other responses.
Aside from \textit{Aarea}, the missingness of each response variable is often explained by its own values.
Specifically, higher values of \textit{SLA} are more likely to be missing, while higher values of \textit{Narea}, \textit{Parea}, and \textit{Gs} are more likely to be observed.
This is consistent with the missBART1 imputations shown in Figure \ref{fig: ypred missBART1and2}.
Despite \textit{Parea} exhibiting the highest level of missingness, only the error bars for \textit{Narea} and \textit{Parea} do not overlap with 0 in the missingness for \textit{Parea}, indicating that its missingness is primarily influenced by itself and \textit{Narea}.
Overall, most of the intervals for $\mathbf{B}_Y$ do not overlap with 0, showing a strong presence of MNAR missingness in the \textit{global Amax} data.
From missBART2, we obtain the variable importance from both the regression and missingness trees.
This is shown in Figure \ref{fig: var imp population pyramid} below.
In Figure \ref{fig: 10 varimp regression}, the importance measures from both sets of trees are displayed for the $10$ most important variables from the regression trees.
While bulk density (\textit{BULK}), calcium carbonate content (\textit{CARB}), and mean monthly fractional sunshine duration (\textit{SUNmin}) were the variables most frequently used in the regression trees for predicting the responses, they showed little importance in explaining the missingness of the data. 
In contrast, Figure \ref{fig: 10 varimp missing} shows that the most influential variables for the missingness model are exchangeable aluminum percentage (\textit{ALU}), clay content (\textit{CLAY}), and the seasonality of precipitation (\textit{PPTseason}). 
Of the $5$ responses variables, \textit{Parea} is the only variable in the top 10 most important variables of the missingness trees, while the others are among the 20 least important variables. 
We also note that \textit{Parea} has no importance in the regression trees, as response variables only contribute to the missingness model by construction.
The covariates \textit{ALU}, relative humidity (\textit{RH}), sand content (\textit{SAND}), and maximum monthly precipitation (\textit{PPTmax}) are among the top 10 most influential variables for predicting the responses as well as explaining the missingness.
\begin{figure}[H]
    \centering
    \begin{subfigure}[b]{0.49\textwidth}
        \centering
        \includegraphics[width=\linewidth]{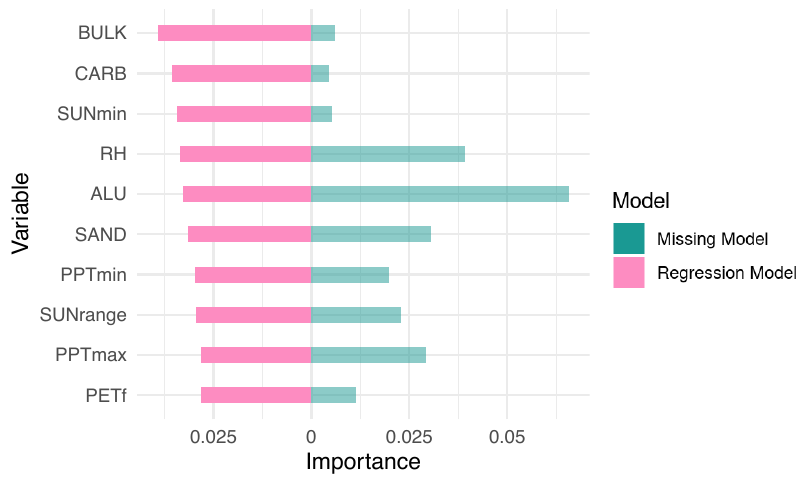}
        \caption{Top 10 variables from the regression trees.}
        \label{fig: 10 varimp regression}
    \end{subfigure}
    \hfill 
    \begin{subfigure}[b]{0.49\textwidth}
        \centering
        \includegraphics[width=\linewidth]{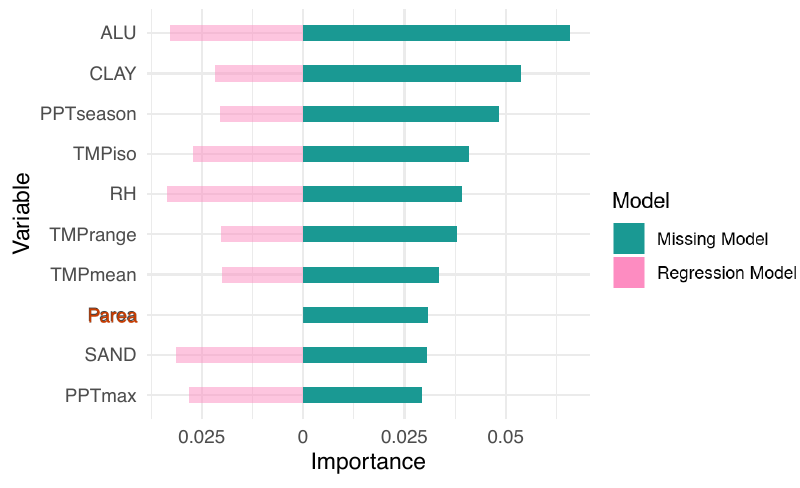}
        \caption{Top 10 variables from the missingness trees.}
        \label{fig: 10 varimp missing}
    \end{subfigure}
    \caption{Comparisons of the top 10 most important variables in the regression and missingness models from missBART2. The importance for each set of trees is given in each case. Only \textit{RH}, \textit{SAND}, and \textit{PPTmax} are common to both. This highlights differences in how variables influence both the missingness mechanism and the regression model's predictions. In (b), \textit{Parea} is the only response variable among the top 10 important variables.
    Response variables have no importance in the regression trees and can only contribute to the missingness model by construction.}
    \label{fig: var imp population pyramid}
\end{figure}
In addition to variable importance, variable interactions in the regression and missingness trees can be measured by counting the number of times two variables appear consecutively along the same branch in the trees.
Using visualization techniques adapted from \citet{inglis2022visualizing}, the average interaction effects of the top 10 variables from the regression and missingness trees are shown using heat maps in Figure \ref{fig: top 10 varimp and varint} below, with variable importance shown on the diagonals and variable interactions on the off-diagonals.
In Figure \ref{fig: 10 varint reg}, \textit{BULK} and potential evapotranspiration (\textit{PETf}) show strong interactions with \textit{SUNmin} implying that \textit{BULK} and \textit{PETf} are often used in conjunction with \textit{SUNmin} for predicting the responses.
In Figure \ref{fig: 10 varint missing}, \textit{PPTmax} and isothermality (\textit{TMPiso}) display the strongest interaction, followed by \textit{CLAY} and mean annual temperature (\textit{TMPmean}).
\begin{figure}[H]
\centering
    \begin{subfigure}[t]{.45\textwidth}
      \centering
      \includegraphics[width=\linewidth]{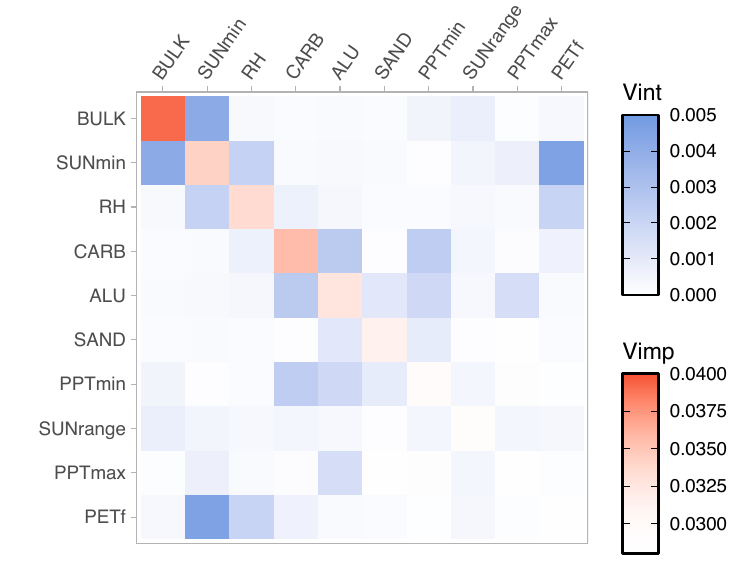}
      \caption{missBART2 regression trees.}
      \label{fig: 10 varint reg}
    \end{subfigure}
    \hfill
    \begin{subfigure}[t]{.45\textwidth}
      \centering
      \includegraphics[width=\linewidth]{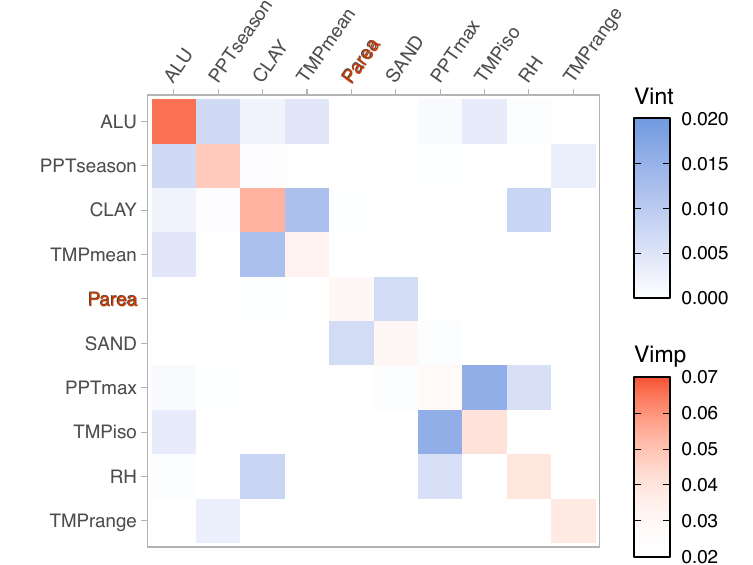}
      \caption{missBART2 missingness trees.}
      \label{fig: 10 varint missing}
    \end{subfigure}
    \caption{Heat maps showing the top 10 important variables and their interactions from the regression trees (a) and missingness trees (b) of missBART2. In (b), \emph{Parea} is highlighted as it is a response variable in the data model and plays the role of a predictor in the missingness model.}
    \label{fig: top 10 varimp and varint}
\end{figure}

\begin{figure}[H]
    \centering
    \includegraphics[width=\textwidth]{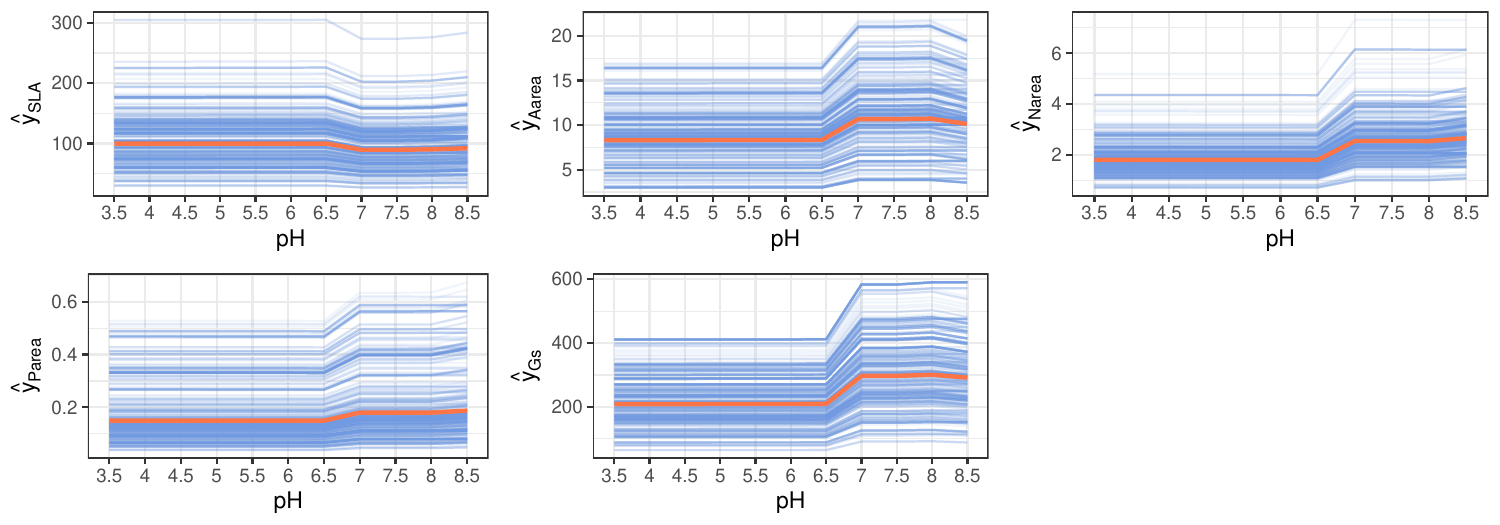}
    \caption{PDP + ICE curves for all responses across different levels of \textit{pH}. \textit{Aarea}, \textit{Narea}, \textit{Parea}, and \textit{Gs} show an increase when \textit{pH} is above 6.5, while \textit{SLA} sees a decline. However, there is a small decrease in \textit{Aarea} as \textit{pH} increases from 8 to 8.5, likewise for \textit{Gs}.}
    \label{fig: pH pdp}
\end{figure}

\begin{figure}[H]
    \centering
    \includegraphics[width=\linewidth]{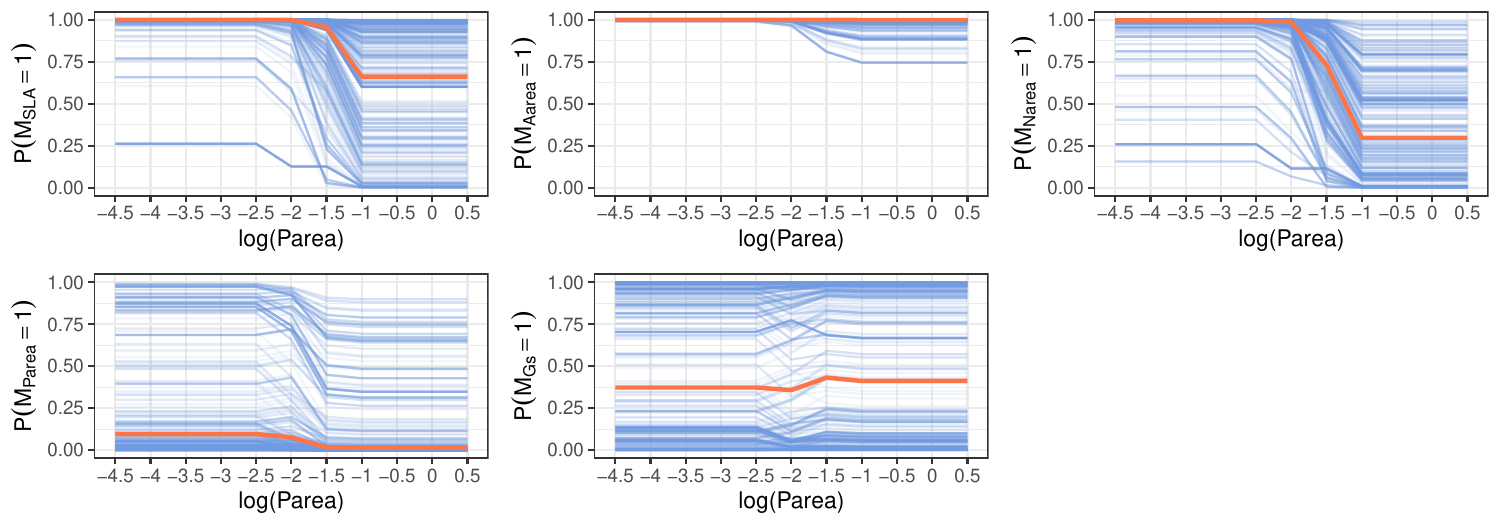}
    \caption{PDP + ICE plots of missingness probabilities for all responses as \textit{Parea} varies. All responses except \textit{Gs} are more likely to be missing when $\log(\textit{Parea})$ exceeds $-2$. }
    \label{fig: PDP missing Parea}
\end{figure}

\citet{maire2015global} found that \textit{Aarea}, \textit{Narea}, and \textit{Parea} increased with higher \textit{pH} and lower \textit{Miq}, \textit{SLA} decreased with \textit{pH}, and \textit{Parea} increased while \textit{Gs} decreased with rising \textit{Pavail}.
Using partial dependence plots \citep[PDP;][]{friedman2001greedy} and individual conditional expectation \citep[ICE;][]{goldstein2015peeking} curves, we further investigate the marginal effect each variable has on the responses.  
In Figure \ref{fig: pH pdp}, \textit{Aarea}, \textit{Narea}, \textit{Parea}, and \textit{Gs} all show an increase when \textit{pH} is above 6.5, while \textit{SLA} sees a decline. However, \textit{Aarea} decreases slightly as \textit{pH} increases from 8 to 8.5, likewise for \textit{Gs}.
These results align with established theory, as \textit{pH} strongly influences plant availability of soil nutrients, directly impacting photosynthetic traits.
More specifically, nutrient availability is generally highest at moderate \textit{pH} levels and decreases significantly in highly alkaline soils (e.g., \textit{pH} above 8) and highly acidic soils (e.g., \textit{pH} below 6) \citep{westerband2023coordination}.
Further analysis shows that as \textit{Miq} decreases, \textit{Aarea} is unaffected, \textit{Narea} and \textit{Parea} increase, and \textit{SLA} and \textit{Gs} decrease.
Apart from a slight decrease in \textit{Gs} when \textit{Pavail} increases, we see virtually no changes in all other responses.
The PDP and ICE plots for \textit{Miq} and \textit{Pavail} are included in \ref{Appendix: PDP Miq Pavail}.
As for the missingness, Figure \ref{fig: PDP missing Parea} shows the effect of $\log(\textit{Parea})$ on the detection probabilities of each response variable. 
While the detection probabilities of most responses decrease as $\log(\textit{Parea})$ exceeds a value of $-2$, the detection probabilities of \textit{Gs} show a slight increase on average.
This implies that \textit{SLA}, \textit{Aarea}, \textit{Narea}, and \textit{Parea} are more likely to be missing when $\log(\textit{Parea})$ is greater than $-2$, while \textit{Gs} shows the opposite.

\section{Discussion \& Conclusion}
\label{sec: 8 Discussion}
Motivated by the \textit{global Amax} data with multivariate missing responses and completely observed covariates, we propose two novel models, missBART1 and missBART2, to address the limitations of existing missing data methods which predominantly focus on MCAR or MAR assumptions. 
Our models, which can handle MCAR, MAR, and MNAR scenarios for univariate and multivariate response data, offer a more flexible approach for predictive modeling of data where missingness may be a concern.
Both models operate within the selection model framework, differing primarily in the specification of the missing data model, where missBART1 uses probit regression and missBART2 uses a probit BART model.
While both models were designed to handle non-ignorable missing responses, they have also been adapted to handle ignorable missingness in the covariates.
In missBART1, prior covariate imputation is necessary before model fitting, while missBART2 directly incorporates covariate missingness within the splitting rules of the decision trees. Both models provide robust alternatives for predictive modeling with partially observed responses, addressing the challenges of MNAR data and offering flexibility for diverse real-world applications.

Our analysis of the \textit{global Amax} data diverges from the approach used by \citet{maire2015global} in several key ways. 
While \citet{maire2015global} employ separate regression models for each photosynthetic trait, potentially overlooking the effects of missingness in the data, our models utilize a non-linear and non-parametric multivariate BART framework, explicitly accounting for the missingness structure in the responses. 
This joint modeling approach allows for a more comprehensive understanding of the relationships between covariates and responses while addressing the limitations of missing data.
In their study, \citet{maire2015global} identified three covariates --- \textit{pH}, \textit{Miq}, and \textit{Pavail} --- as significantly influencing photosynthetic traits. 
While our PDP + ICE plots generally support these findings for \textit{pH} and \textit{Miq}, our models suggest little to no relationship between \textit{Pavail} and the responses. 
Furthermore, the univariate analyses in \citet{maire2015global} found that most responses were strongly influenced by \textit{pH} and \textit{Miq}, whereas \textit{Pavail} primarily affected \textit{Parea} and \textit{Gs} --- the two responses with the highest proportions of missingness. 
This raises two important considerations. First, explicitly modeling the missingness structure allows our analysis to uncover more nuanced relationships between the covariates and responses, highlighting potential biases that may arise when missing data mechanisms are ignored. 
Second, future modification to our models could involve integrating the ``seemingly unrelated BART" model from \citet{esser2024seemingly}, allowing each response variable to be associated with different sets of BART trees while also accounting for dependencies between the responses, incorporating  further flexibility and interpretability in multivariate response settings.

\section*{Acknowledgements}
Yong Chen Goh was supported by INSIGHT Phase 2 SFI/12/RC/2289\_P2. 
Andrew Parnell’s work was supported by: an SFI Research Centre award 12/RC/2289\_P2; a Science Foundation Ireland Career Development Award 17/CDA/4695; an investigator award (16/IA/4520); a Marine Research Programme funded by the Irish Government, co-financed by the European Regional Development Fund (Grant-Aid Agreement No. PBA/CC/18/01); European Union’s Horizon 2020 research and innovation programme (grant agreement No. 818144); SFI Centre for Research Training 18CRT/6049; and Research Ireland Co-Centre award 22/CC/11103. For the purpose of Open Access, the author has applied a CC BY public copyright licence to any Author Accepted Manuscript version arising from this submission. We also thank Dr.~Alan Inglis for helping with the variable importance and interactions heat maps in Section \ref{sec: 7 Real data example}.

\bibliographystyle{plainnat}
\bibliography{references}

\newpage
\appendix
\numberwithin{figure}{section}
\numberwithin{table}{section}
\renewcommand{\thesection}{Appendix \Alph{section}}
\renewcommand{\thefigure}{\Alph{section}.\arabic{figure}}
\renewcommand{\thetable}{\Alph{section}.\arabic{table}}

\section{Schematic diagram of joint models}
\label{Appendix: schematic diagram}
\begin{figure}[H]
    \centering
    \includegraphics[width=\textwidth]{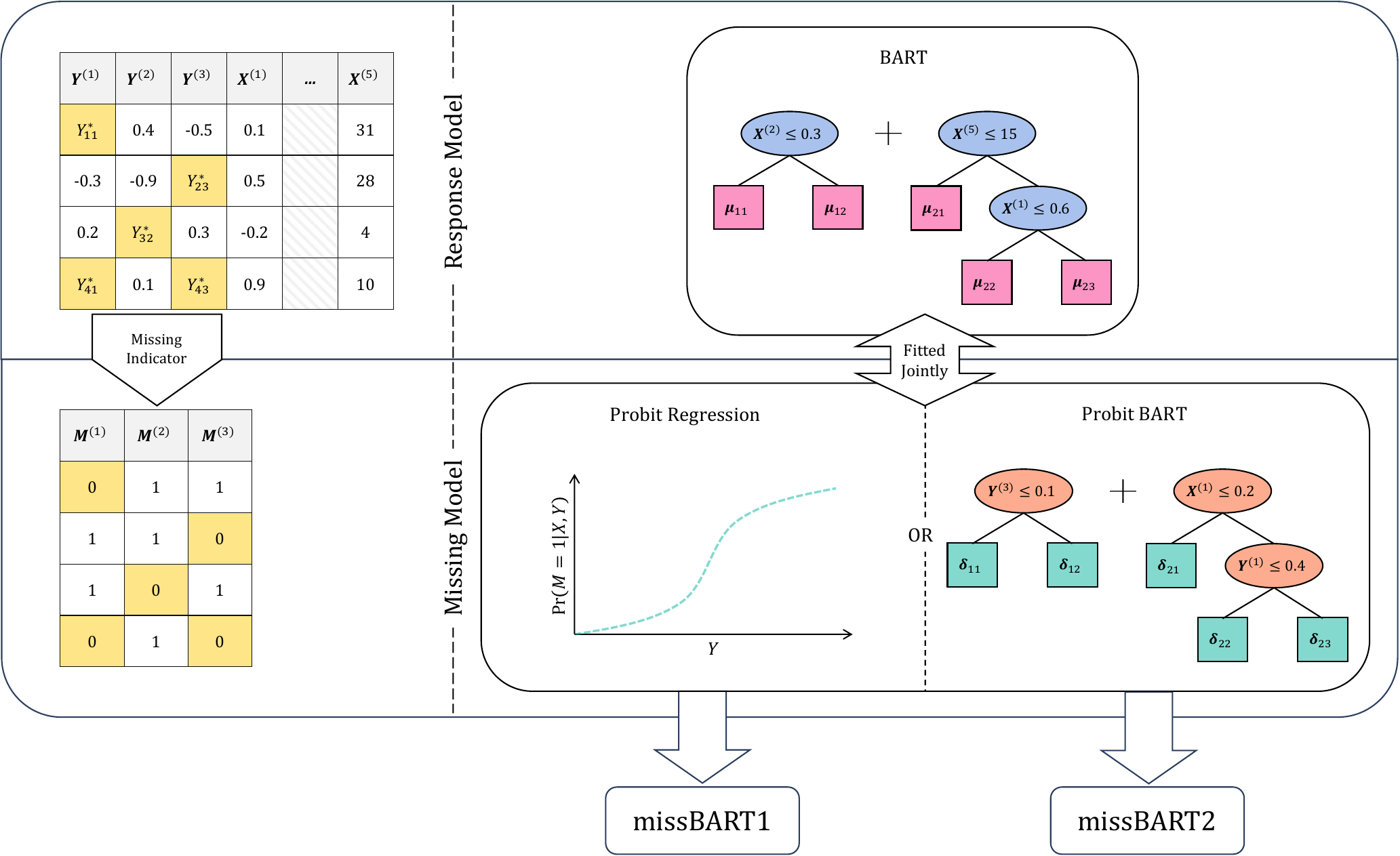}
    \caption{Schematic diagram of missBART1 and missBART2 using a toy dataset with three response variables $\left(\mathbf{Y}^{(1)},\ldots,\mathbf{Y}^{(3)}\right)$ and five covariates $\left(\mathbf{X}^{(1)},\ldots,\mathbf{X}^{(5)}\right)$. The responses have missing entries, denoted by $Y_{ij}^\ast$ with row index $i$ and column index $j$. Each $\mathbf{M}^{(j)}$ is the resulting missing data indicator for $\mathbf{Y}^{(j)}$, where $M_{ij}=0$ if $Y_{ij}$ is missing and \emph{vice versa}. 
    Both joint models fit a BART model to the responses, using $\mathbf{X}$ in the splitting rules of the trees.
    In missBART1, a probit regression model is jointly fitted to model the missing data indicators $\mathbf{M}$, where $\mathbf{X}$ and $\mathbf{Y}$ are used as the missing model covariates.
    In missBART2, a probit BART model is fitted to $\mathbf{M}$, with $\mathbf{X}$ and $\mathbf{Y}$  used in the splitting rules of the trees. 
    While the goal is to account for unobserved responses, both models are also capable of handling covariate missingness, either via prior covariate imputation (missBART1) or incorporating missingness into the splitting rules of the trees (missBART2). 
    }
    \label{fig: Joint model diagram}
\end{figure}

A schematic diagram of missBART1 and missBART2 using a toy dataset is depicted in Figure \ref{fig: Joint model diagram}. 

\section{Posterior Distribution of \texorpdfstring{$\boldsymbol{\Psi}$}{Psi}}
\label{Appendix: posterior dist of Psi}
We derive the posterior distribution of $\boldsymbol{\Psi}$ where $\boldsymbol{\Psi}^{-1} = \operatorname{diag}(\tau_{B_0}, \tau_{B_{X}} \boldsymbol{1}_q, \tau_{B_{Y}}\boldsymbol{1}_p)$.
Given the following priors
\begin{align*}
    \mathbf{B}\given \boldsymbol{\Psi}, \mathbf{R} &\sim \mathcal{MN}_{r \times p} (\mathbf{0}, \boldsymbol{\Psi}, \mathbf{R})\\
    \tau_{B_0} &\sim Ga(\alpha_0, \beta_0)\\
    \tau_{B_{X}} &\sim Ga(\alpha_X, \beta_X) \\
    \tau_{B_{Y}} &\sim Ga(\alpha_Y, \beta_Y),
\end{align*}
we obtain the following full conditional distribution
\begin{align*}
    p(\tau_{B_0}, \tau_{B_{X}}, \tau_{B_{Y}} \given \cdot) &\propto \lvert\boldsymbol{\Psi}\rvert^{-\frac{p}{2}} \exp \left\{-\frac{1}{2} \operatorname{tr}(\mathbf{R}^{-1} \mathbf{B}^\top \boldsymbol{\Psi}^{-1} \mathbf{B})\right\} \times \tau_{B_0}^{\alpha_0 - 1} \exp\left\{-\beta_0 \tau_{B_0}\right\} \\
    &\quad  \times \tau_{B_{X}}^{\alpha_X - 1} \exp\left\{-\beta_X \tau_{B_{X}} \right\} \times \tau_{B_{Y}}^{\alpha_Y - 1} \exp\left\{-\beta_Y \tau_{B_{Y}} \right\}.
\end{align*}
By using the trace property 
$$
    \operatorname{tr}(\mathbf{R}^{-1} \mathbf{B}^\top \boldsymbol{\Psi}^{-1} \mathbf{B}) = \operatorname{tr}(\boldsymbol{\Psi}^{-1} \mathbf{B} \mathbf{R}^{-1} \mathbf{B}^\top)
$$
and letting $\mathbf{A} = \mathbf{B} \mathbf{R}^{-1} \mathbf{B}^\top$, we obtain
\begin{align*}
    \operatorname{tr}(\boldsymbol{\Psi}^{-1} \mathbf{A}) &= \operatorname{tr}\begin{pmatrix}
    \tau_{B_0} A_1 \\ \tau_{B_{X}} A_2 \\ \ldots \\ \tau_{B_{X}} A_{1+q} \\ \tau_{B_{Y}} A_{2+q} \\ \ldots \\ \tau_{B_{Y}} A_r
\end{pmatrix} \\
&= \tau_{B_0} A_{11} + \tau_{B_{X}} A_{22} + \ldots + \tau_{B_{X}}A_{(1+q)(1+q)} + \tau_{B_{Y}}A_{(2+q)(2+q)} + \ldots + \tau_{B_{Y}} A_{rr},
\end{align*}
where $A_i, i=1,\ldots,r$, is the $i^{th}$ row of the matrix $\mathbf{A}$, and $A_{ii}$ is the $i^{th}$ diagonal entry of $\mathbf{A}$. From here, we have
\begin{align*}
    p(\tau_{B_0}, \tau_{B_{X}}, \tau_{B_{Y}} \given \cdot) \propto & \left(\tau_{B_0} \times \tau_{B_{X}}^{q} \times \tau_{B_{Y}}^{p} \right)^{\frac{p}{2}} \exp \left\{-\frac{1}{2} \left(\tau_{B_0}A_{11} + \tau_{B_{X}} \sum_{i=2}^{1+q} A_{ii} + \tau_{B_{Y}} \sum_{i=2+q}^{r} A_{ii} \right)\right\} \\
    & \times \tau_{B_0}^{\alpha_0 - 1} \exp\left\{-\beta_0 \tau_{B_0}\right\} \times \tau_{B_{X}}^{\alpha_X - 1} \exp\left\{-\beta_X \tau_{B_{X}}\right\} \times \tau_{B_{Y}}^{\alpha_Y - 1} \exp\left\{-\beta_Y \tau_{B_{Y}}\right\},
\end{align*}
from which we can easily see that
\begin{align*}
    \tau_{B_0}\given\cdot &\propto \tau_{B_0}^\frac{p}{2} \exp \left\{-\frac{1}{2} \tau_{B_0} A_{11} \right\} \tau_{B_0}^{\alpha_0 - 1} \exp \left\{-\beta_0 \tau_{B_0} \right\} \\
    &\propto \tau_{B_0}^{\frac{p}{2} + \alpha_0 -1} \exp \left\{ -\left(\frac{1}{2}A_{11} + \beta_0 \right) \tau_{B_0} \right\} \\
    \tau_{B_{X}}\given\cdot &\propto \tau_{B_{X}}^\frac{pq}{2} \exp \left\{-\frac{1}{2} \tau_{B_{X}} \sum_{i=2}^{1+q} A_{ii} \right\}\tau_{B_{X}}^{\alpha_X - 1} \exp\left\{-\beta_X \tau_{B_{X}} \right\} \\
    &\propto \tau_{B_{X}}^{\frac{pq}{2} + \alpha_X - 1} \exp \left\{-\left(\frac{1}{2} \sum_{i=2}^{1+q} A_{ii} + \beta_X \right)\tau_{B_{X}} \right\} \\
    \tau_{B_{Y}}\given\cdot &\propto \tau_{B_{Y}}^\frac{p^2}{2} \exp \left\{-\frac{1}{2} \tau_{B_{Y}} \sum_{i=2+q}^{r} A_{ii} \right\}\tau_{B_{Y}}^{\alpha_Y - 1} \exp\left\{-\beta_Y \tau_{B_{Y}} \right\} \\
    &\propto \tau_{B_{Y}}^{\frac{p^2}{2} + \alpha_Y - 1} \exp \left\{-\left(\frac{1}{2} \sum_{i=2+q}^{r} A_{ii} + \beta_Y \right)\tau_{B_{Y}} \right\}.
\end{align*}
Finally, the posterior distributions are
\begin{align*}
    \tau_{B_0}\given\cdot &\sim Ga \left(\frac{p}{2} + \alpha_0, \frac{A_{11}}{2}+ \beta_0 \right) \\
    \tau_{B_{X}}\given\cdot &\sim Ga \left(\frac{pq}{2} + \alpha_X, \frac{\sum_{i=2}^{1+q} A_{ii}}{2} + \beta_X \right)\\
    \tau_{B_{Y}}\given\cdot &\sim Ga \left(\frac{p^2}{2} + \alpha_Y, \frac{\sum_{i=2+q}^{r} A_{ii}}{2} + \beta_Y \right).
\end{align*}

\section{Sampling algorithms for joint models} 
\label{Appendix: posterior sampling missBART1&2}
We describe the posterior sampling algorithms in detail, for missBART1 and missBART2, in \ref{Appendix: sampling missBART1} and  \ref{Appendix: sampling missBART2}, respectively. 
\subsection{missBART1 sampling algorithm} \label{Appendix: sampling missBART1}
The posterior sampling algorithm for missBART1 is outlined here:
\begin{enumerate}
    \item For all $K$ trees, propose a new tree via a grow, prune, change, or swap move (see \citet{kapelner2016bartmachine} for details on these tree-proposal moves) and accept or reject using a Metropolis-Hastings step. 
    For notational simplicity, we drop the tree index $k$ and thus the tree posterior takes the form
    \begin{equation*}
        p\left(\mathcal{T}\given \mathbf{r}, \boldsymbol{\Omega}\right) \propto \pi\left(\mathcal{T}\right) \prod_{\ell} p\left(\mathbf{r}_\ell\given\mathcal{T}, \boldsymbol{\Omega}\right),
    \end{equation*}
    where $\pi(\mathcal{T})$ denotes the tree prior from \citet{chipman2010bart} and $\mathbf{r}_\ell$ denotes the partial residuals 
    \begin{equation*}
        \mathbf{r}_k \equiv \mathbf{Y} - \sum_{t \neq k}g\left(\mathbf{X}; \mathcal{T}_t, \mathbf{Q}_t\right)
    \end{equation*}
    assigned to terminal node $\ell$ in tree $k$. Note that
    \begin{align*}
        p\left(\mathbf{r}_\ell\given\mathcal{T}, \boldsymbol{\Omega}\right) &= \int p\left(\mathbf{r}_\ell\given\mathcal{T}, \boldsymbol{\mu}_\ell, \boldsymbol{\Omega}\right) \pi\left(\boldsymbol{\mu}_\ell\right)\dd\boldsymbol{\mu}_\ell \\
        &= \left(2\pi\right)^{-\frac{n_\ell p}{2}} \tau_\mu ^{\frac{p}{2}}  \lvert\boldsymbol{\Omega}\rvert^{\frac{n_\ell}{2}} \lvert\boldsymbol{\Sigma}_{\mu}\rvert^{\frac{1}{2}} \exp \left\{-\frac{1}{2} \left\lbrack \boldsymbol{\mu}_0^\top \left(\tau_\mu \mathbf{I}_p\right) \boldsymbol{\mu}_0 - \boldsymbol{\mu}_r^\top \boldsymbol{\Sigma}_r^{-1} \boldsymbol{\mu}_r + \sum_{i=1}^{n_\ell} \mathbf{r}_{\ell i}^\top \boldsymbol{\Omega} \mathbf{r}_{\ell i} \right\rbrack \right\},
    \end{align*}
    where $\boldsymbol{\mu}_r = \boldsymbol{\Sigma}_r \left\lbrack\boldsymbol{\Omega}(\sum_{i=1}^{n_\ell} \mathbf{r}_{\ell i}) + (\tau_\mu \mathbf{I}_p) \boldsymbol{\mu}_0 \right\rbrack$, $\boldsymbol{\Sigma}_r^{-1} = n_\ell \boldsymbol{\Omega} + \tau_\mu \mathbf{I}_p$, and $n_\ell$ denotes the total number of observations which fall under terminal node $\ell$.
    
    \item For each terminal node $\ell$ in tree $k$, once again dropping the tree index $k$, make a draw of $\boldsymbol{\mu}_{\ell} \in \mathbf{Q}$ from $\boldsymbol{\mu}_{\ell}\given\mathbf{r}_\ell, \mathcal{T}, \boldsymbol{\Omega} \sim \mathcal{N}_p\left(\boldsymbol{\mu}_\mu, \boldsymbol{\Sigma}_{\mu}\right)$
    where $\boldsymbol{\mu}_\mu = \boldsymbol{\Sigma}_{\mu} \boldsymbol{\Omega} \sum_{i=1}^{n_\ell} \mathbf{r}_{\ell i}$ and $\boldsymbol{\Sigma}_{\mu} = \left(n_\ell \boldsymbol{\Omega} + \tau_\mu \mathbf{I}_p \right)^{-1}$.
    
    \item After carrying out steps $1$ and $2$ for all $K$ trees, sample $\boldsymbol{\Omega}$ from $\boldsymbol{\Omega}\given\mathcal{T},\mathbf{Q},\Tilde{\mathbf{Y}} \sim \mathcal{W}_p \left(n+\nu, \mathbf{V}_{\Omega} \right)$
    where $\mathbf{V}_{\Omega}^{-1} = \sum_{i=1}^n\left(\Tilde{\mathbf{Y}}_i - \widehat{\Tilde{\mathbf{Y}}}_i \right) \left(\Tilde{\mathbf{Y}}_i - \widehat{\Tilde{\mathbf{Y}}}_i \right)^\top + \mathbf{V}^{-1}$ and
    
    $\widehat{\Tilde{\mathbf{Y}}}_i=\sum_{k=1}^K g\left(\mathbf{X}_i;\mathcal{T}_k,\mathbf{Q}_k\right)$.

    \item The posterior distribution of $\mathbf{M}^\star_{i}$ follows a multivariate truncated normal distribution \citep{damien2001sampling}
    $\mathbf{M}_i^\star \given \mathbf{X}_i, \Tilde{\mathbf{Y}}_i, \mathbf{B}, \mathbf{M}_i \sim \mathcal{TN}_p(\mathbf{B}^\top\mathbf{Z}_i,\mathbf{R},\boldsymbol{\gamma}_i)$
    where $\mathbf{Z}_i = (1, \mathbf{X}_i, \Tilde{\mathbf{Y}}_i)^\top$ and $\boldsymbol{\gamma}_i$ denotes the $p$-dimensional vector of truncation points such that $\gamma_{ij}=[0,\infty)$ if $M_{ij}=1$ and $\gamma_{ij}=(-\infty,0]$ if $M_{ij}=0$.
    
    \item To sample $\mathbf{Y}^{mis}$, we make draws from 
        \begin{equation*}
        \mathbf{Y}^{mis}_i \given \mathbf{X}, \mathbf{Y}^{obs}_i, \bm{\mathcal{T}},\mathbf{Q},\boldsymbol{\Omega}, \mathbf{M}^\star_{i}, \mathbf{B}, \mathbf{R} \sim \mathcal{N}_{p_i} \left(\left\lbrack\boldsymbol{\mu}_{Y}\right\rbrack_{\mathcal{M}_i}, [\boldsymbol{\Sigma}_{Y}]_{\mathcal{M}_i, \mathcal{M}_i}\right),
    \end{equation*}
    where $\mathcal{M}_i = \{j \given M_{ij}=0 \}$ is the set of column indices where $\Tilde{\mathbf{Y}}_i$ is missing,
    $p_i$ is equal to the number of elements in $\mathcal{M}_i$,
    $\boldsymbol{\Sigma}_{Y} = \left(\boldsymbol{\Omega} + \mathbf{B}_Y \mathbf{R}^{-1} \mathbf{B}_Y^\top \right)^{-1}$,
    $\boldsymbol{\mu}_{Y} = \boldsymbol{\Sigma}_{Y} \left\lbrack\boldsymbol{\Omega}\widehat{\Tilde{\mathbf{Y}}} _i + \mathbf{B}_{Y} \mathbf{R}^{-1} \left(\mathbf{M}^\star_{i} - \mathbf{B}^\top_{(Y)} \mathbf{Z}_{(Y)i} \right)\right\rbrack$, 
    $\left\lbrack\boldsymbol{\mu}_{Y}\right\rbrack_{\mathcal{M}_i}$ is the $p_i$-dimensional subset of $\boldsymbol{\mu}_{Y}$ obtained by extracting the $\mathcal{M}_i$ elements from $\boldsymbol{\mu}_{Y}$, and 
    $[\boldsymbol{\Sigma}_{Y}]_{\mathcal{M}_i, \mathcal{M}_i}$ is the $p_i \times p_i$ submatrix of $\boldsymbol{\Sigma}_{Y}$ obtained in a similar fashion.
    Additionally, 
    \begin{equation*}
        \mathbf{B} = \begin{bmatrix}\mathbf{B}_{(Y)}\\\mathbf{B}_Y\end{bmatrix} =
        \begin{bmatrix}\mathbf{B}_{0}\\\mathbf{B}_{X}\\\mathbf{B}_Y\end{bmatrix} =
        \begin{bmatrix} 
            b_{11} & \ldots & b_{1p} \\
            b_{21} & \ldots & b_{2p} \\
            \vdots & \ldots  & \vdots \\
            b_{(1+q)1} & \ldots & b_{(1+q)p}\\
            b_{(r-p+1)1} & \ldots & b_{(r-p+1)p}\\
            \vdots & \ldots & \vdots\\
            b_{r1} & \ldots & b_{rp}
        \end{bmatrix}.
    \end{equation*}
    $\mathbf{B}_{Y}$ is the $p \times p$ submatrix of $\mathbf{B}$ obtained by removing its first $1+q$ rows, and $\mathbf{B}_{(Y)}=\{\mathbf{B}_0,\mathbf{B}_X\}$ is the complementary $(1+q) \times p$ submatrix with probit regression parameters associated with the intercept and covariates $\mathbf{X}$. $\mathbf{Z}_{(Y)i}$ denotes the subset of predictors comprising only the intercept term and $\mathbf{X}_i$, i.e. $\mathbf{Z}_{(Y)i} = (1, X_{i1}, \ldots, X_{iq})^\top$. 
    
    \item Sample ($\mathbf{B}$, $\mathbf{R}$) as in \citet{talhouk2012efficient}.

    \item Sample $\boldsymbol{\Psi}^{-1} = \operatorname{diag}(\tau_{B_0}, \tau_{B_{X}} \boldsymbol{1}_q, \tau_{B_{Y}}\boldsymbol{1}_p)$ from
    \begin{align*}
        \tau_{B_0}\given\cdot &\sim Ga \left(\frac{p}{2} + \alpha_0, \frac{A_{11}}{2}+ \beta_0 \right) \\
        \tau_{B_{X}}\given\cdot &\sim Ga \left(\frac{pq}{2} + \alpha_X, \frac{\sum_{i=2}^{1+q} A_{ii}}{2} + \beta_X \right)\\
        \tau_{B_{Y}}\given\cdot &\sim Ga \left(\frac{p^2}{2} + \alpha_Y, \frac{\sum_{i=2+q}^{r} A_{ii}}{2} + \beta_Y \right),
    \end{align*}
    where $A_{ii}$ is the $i^{th}$ diagonal entry of $\mathbf{A} = \mathbf{B}\mathbf{R}^{-1}\mathbf{B}^\top$.
\end{enumerate}

\subsection{missBART2 sampling algorithm} \label{Appendix: sampling missBART2}
The posterior sampling algorithm for missBART2 is similar, with a few exceptions:
\begin{enumerate}
    \item Repeat Steps 1 and 2 from \ref{Appendix: sampling missBART1} for all $K_y$ trees.

    \item Repeat Steps 1 and 2 from \ref{Appendix: sampling missBART1} for all $K_m$ trees.

    \item Repeat Step 3 from \ref{Appendix: sampling missBART1} for sampling $\boldsymbol{\Omega}$.

    \item The posterior distribution of $\mathbf{M}^\star_{i}$ again follows a multivariate truncated normal distribution such that
    $\mathbf{M}^\star_{i}\given\mathbf{X}_i, \mathbf{Y}_i, \bm{\mathcal{T}}^{m}, \mathbf{Q}^{m}, \mathbf{M}_i \sim \mathcal{TN}_p(\hat{\mathbf{M}}^\star_i, \mathbf{I}_p,\boldsymbol{\gamma}_i)$
    where 
    
    $\hat{\mathbf{M}}^\star_i = \sum_{k_m=1}^{K_m} g(\mathbf{X}_i, \mathbf{Y}_i; T_{k_m}^m, \mathbf{Q}_{k_m}^m)$ and $\boldsymbol{\gamma}_i$ is as defined in Step 4 of \ref{Appendix: sampling missBART1}.

    \item For every missing entry, first propose a new value $Y^{mis}_{t,i,j}$ from $\mathcal{N}(Y^{mis}_{t-1,i,j}, \sigma_Y^2)$. Next, calculate the acceptance probability 
    \begin{equation*}
    \omega(\mathbf{Y}_{t,i}^{mis}, \mathbf{Y}_{t-1,i}^{mis}) =
        \frac{
            p\left(\mathbf{Y}_{t,i}^{mis} \given \mathbf{Y}_{i}^{obs}, \boldsymbol{\theta}\right) p\left(\mathbf{M}_i^\star \given \mathbf{Y}_{i}^{obs}, \mathbf{Y}_{t,i}^{mis}, \boldsymbol{\psi}\right) q(\mathbf{Y}_{t,i}^{mis} \rightarrow \mathbf{Y}_{t-1,i}^{mis})
        }{
            p\left(\mathbf{Y}_{t-1,i}^{mis} \given \mathbf{Y}_{i}^{obs}, \boldsymbol{\theta}\right) p\left(\mathbf{M}_i^\star \given \mathbf{Y}_{i}^{obs}, \mathbf{Y}_{t-1,i}^{mis}, \boldsymbol{\psi}\right) q(\mathbf{Y}_{t-1,i}^{mis} \rightarrow \mathbf{Y}_{t,i}^{mis})
        }
    \end{equation*}
    and accept or reject the proposed $\mathbf{Y}_t^{mis}$ with probability $\text{min}\left(1, \omega(\mathbf{Y}_{t,i}^{mis}, \mathbf{Y}_{t-1,i}^{mis})\right)$.
\end{enumerate}

\section{Univariate simulated examples}
\label{Appendix: univariate examples}
\begin{figure}[H]
     \centering
     \begin{subfigure}[t]{0.49\textwidth}
        \centering
        \includegraphics[width=\linewidth]{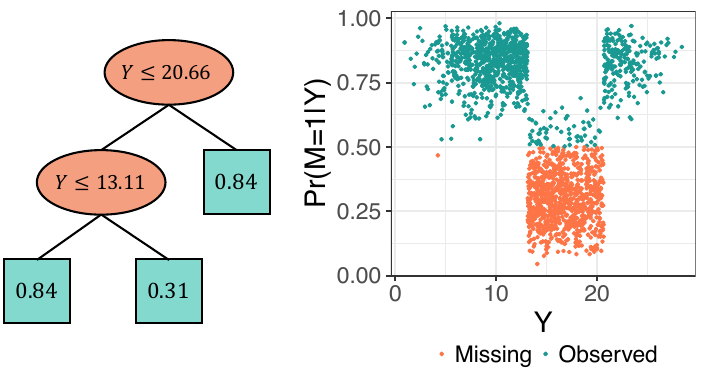}
        \caption{u-shape: Most of the data between $\lbrack13.11, 20.66\rbrack$ are missing.}
        \label{subfig: u-shape tree}
    \end{subfigure}
    \hfill
    \begin{subfigure}[t]{0.49\textwidth}
        \centering
        \includegraphics[width=\linewidth]{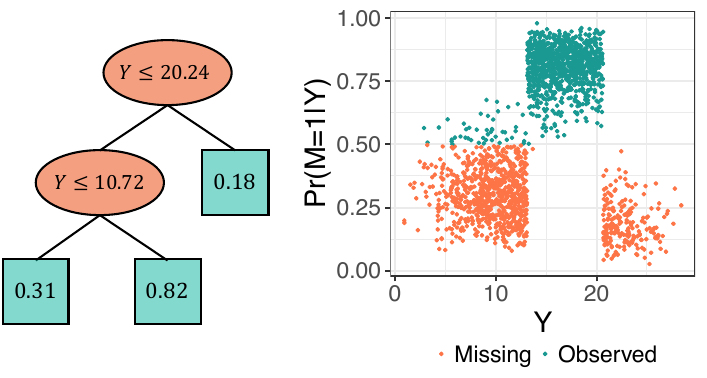}
        \caption{n-shape: Most of the data outside the range $\lbrack10.72, 20.24\rbrack$ are missing.}
        \label{subfig: n-shape tree}
    \end{subfigure}
    \caption{Missingness trees with detection probabilities in the terminal nodes and plots of detection probabilities against true $\mathbf{Y}$ values for the u-shape (left) and n-shape (right) examples.}
    \label{fig: non-linear trees}
\end{figure}

\begin{figure}[H]
     \centering
     \begin{subfigure}[b]{\textwidth}
         \centering
         \includegraphics[width=\textwidth]{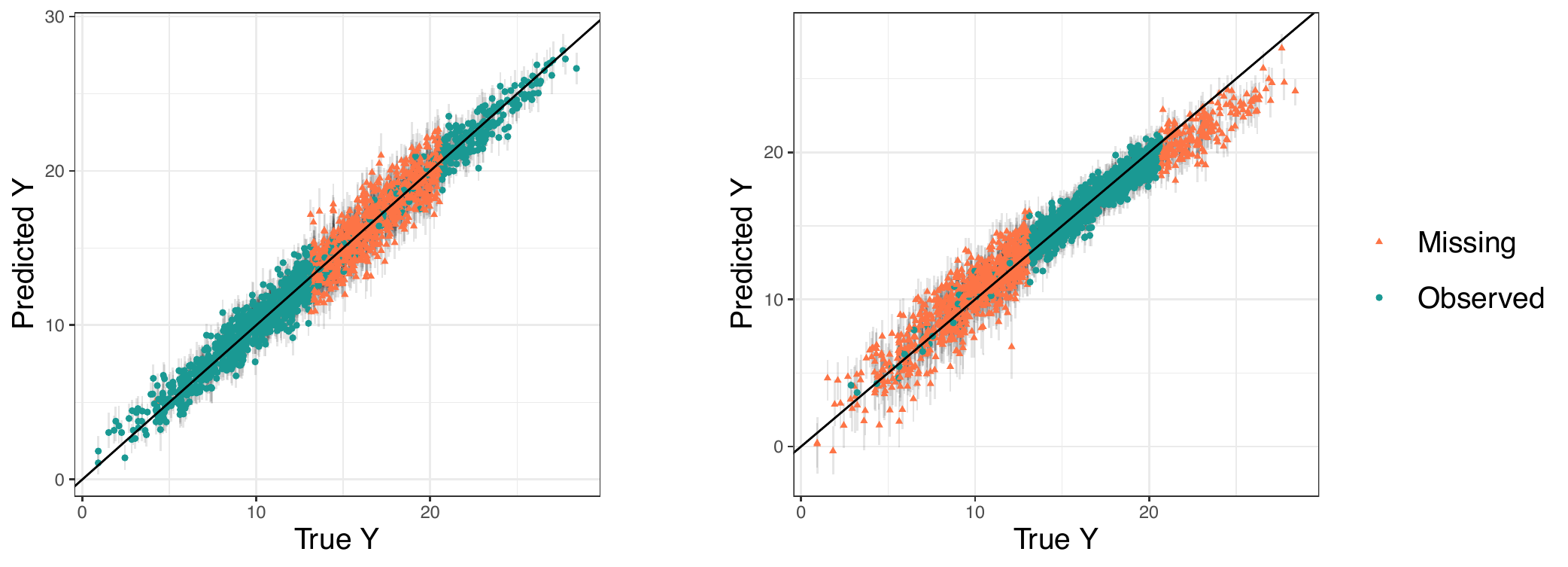}
         \caption{missBART1: the model struggles to capture the upper end of the missing data in the n-shape scenario.}
         \label{subfig: non-linear missBART1}
     \end{subfigure}
     \begin{subfigure}[b]{\textwidth}
         \centering
         \includegraphics[width=\textwidth]{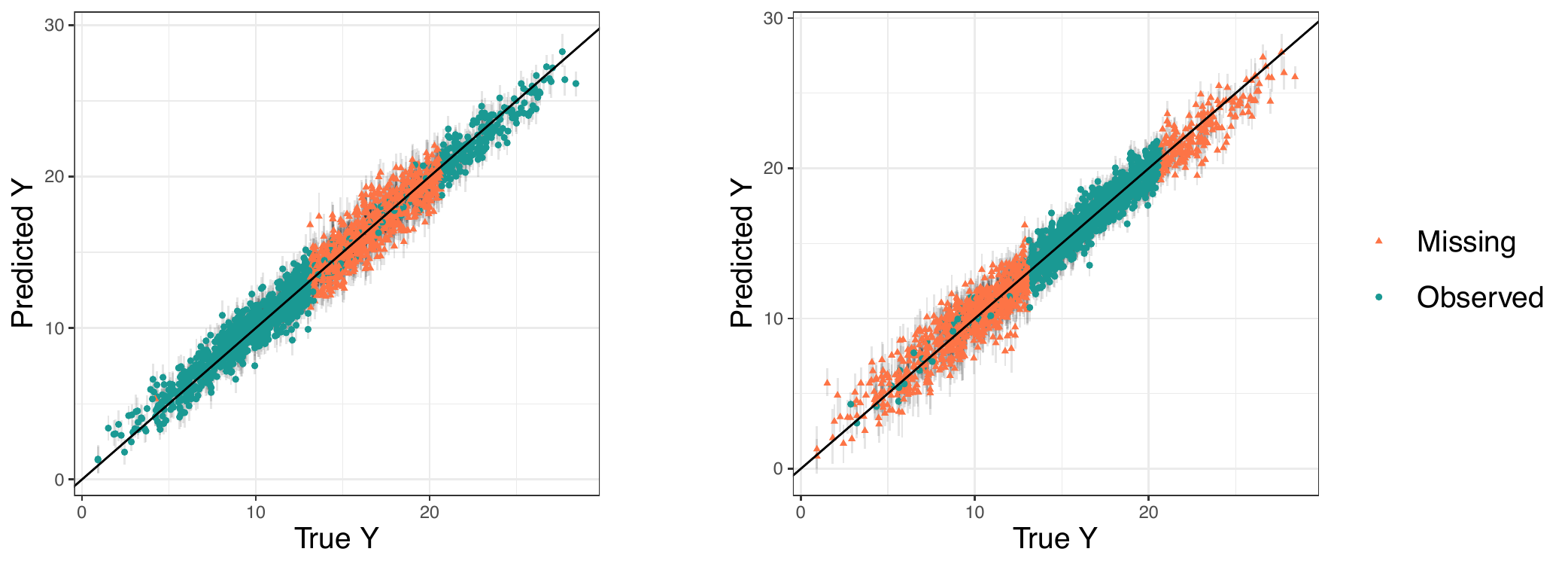}
         \caption{missBART2: the model performs well in both scenarios, closely matching the true simulated values.}
         \label{subfig: non-linear missBART2}
     \end{subfigure}
     \caption{Out-of-sample predictions and posterior imputations from missBART1 (a) and missBART2 (b). The left subfigures show performance on the u-shape missingness pattern, while the right subfigures show performance on the n-shape pattern.  Vertical lines depict the $95\%$ prediction and posterior intervals.}
     \label{fig: non-linear predictions}
\end{figure}

We illustrate two univariate response cases featuring non-linear patterns of MNAR missingness where the detection probabilities of the response vary non-linearly with the values of the responses themselves.
We refer to these two examples as the `u-shape' and `n-shape' missingness.
In both examples, we first generate a complete dataset with $n=2000$ i.i.d samples using the Friedman function \citep{friedman1991multivariate}
\begin{equation}
    \label{eq: Friedman function}
    Y_i = 10 \sin{(\pi X_{i1} X_{i2})} + 20 (X_{i3} - 0.5)^2 + 10 X_{i4} + 5 X_{i5} + \epsilon_i, \quad \epsilon_i \overset{i.i.d}{\sim} N(0,1),
\end{equation}
where each covariate is independently drawn from a $\text{Unif}(0,1)$ distribution. 
Missingness is then induced through a single tree.
Figure \ref{fig: non-linear trees} shows these tree structures and their resulting simulated detection probabilities, $\text{Pr}(M_i=1 \given Y_i)$, against the true $\mathbf{Y}$ values.
All observations with detection probability below $0.5$ are designated as missing and \emph{vice versa}. In the u-shape example, $55.1\%$ of the data are observed, with missing $\mathbf{Y}$ values occurring more frequently in the mid-range of the data. 
In the n-shape example, $50.2\%$ of the data are observed and missing values mostly occur at the extreme ends of the data range.
We run missBART1 and missBART2 for $5000$ burn-in and $5000$ post-burn-in iterations. 
The number of trees in the data model is fixed at $100$, and $20$ probit BART trees are used in missBART2.
Figure \ref{fig: non-linear predictions} shows the out-of-sample predictions for the observed cases and posterior imputations for the missing cases against their true simulated values obtained from both models, along with their 95\% prediction and posterior intervals.
In the u-shape scenario, both models perform well at fitting the observed values as well as making accurate posterior imputations for the missing values. 
However, in the n-shape scenario, while missBART2 maintains accuracy, missBART1 struggles to capture the tail ends of the missing data, particularly in the upper tail where no values are observed. 
This underlines the limitations of missBART1, which assumes a probit missingness pattern, in capturing complex non-linear structures while also highlighting the strong capabilities of missBART2 in capturing non-linear patterns.

\begin{figure}[H]
     \centering
     \begin{subfigure}[t]{0.49\textwidth}
         \centering
         \includegraphics[width=\linewidth]{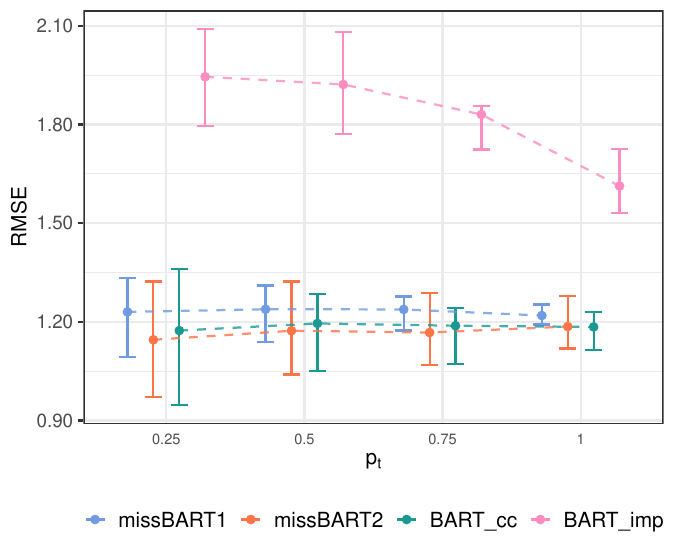}
         \caption{Out-of-sample RMSE for u-shaped missingness.}
         \label{subfig: u-shape RMSE}
     \end{subfigure}
     \hfill
     \begin{subfigure}[t]{0.49\textwidth}
         \centering
         \includegraphics[width=\linewidth]{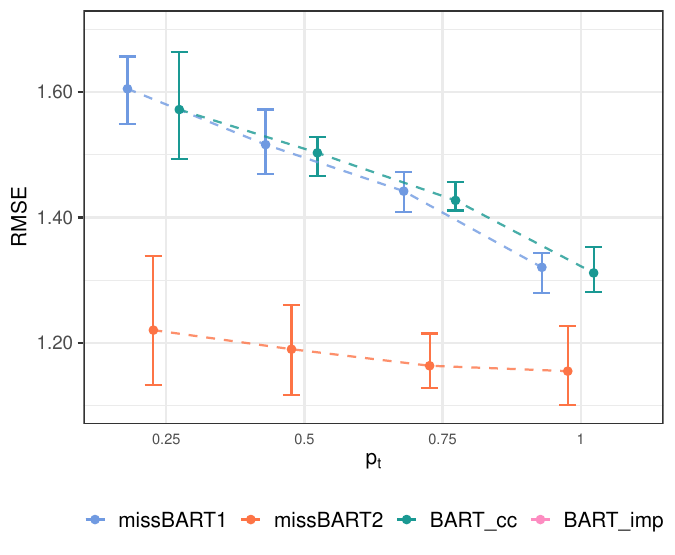}
         \caption{Out-of-sample RMSE for n-shaped missingness.}
         \label{subfig: n-shape RMSE}
     \end{subfigure}
     \caption{Out-of-sample RMSE values from 4-fold cross-validation for data with detection probabilities below thresholds $p_t$. Results for BART\_imp are excluded from (b) due to poor performance.}
     \label{fig: non-linear RMSE}
\end{figure}

Next, through $4$-fold cross-validation, we compare the performances of our joint models with two other approaches: a standard BART model applied to complete cases (`BART\_cc') and applying a standard BART model on the dataset imputed by the \texttt{missForest} package in \texttt{R} \citep{stekhoven2012missforest} (`BART\_imp').
We calculate the out-of-sample RMSEs for observations based on varying levels of detection probability thresholds, $p_t$, i.e. $\text{Pr}(M_i=1 \given Y_i) \leq p_t$, where $p_t=\{0.25, 0.5, 0.75, 1\}$. 
The results are shown in Figure \ref{fig: non-linear RMSE}.
From Figure \ref{subfig: u-shape RMSE}, we note that BART\_imp, which imputes the data using \texttt{missForest} prior to fitting a BART model, consistently underperforms across all detection probability thresholds in both scenarios. 
In Figure \ref{subfig: n-shape RMSE}, the RMSE values from BART\_imp are omitted due to poor performance (between $2.39$ and $3.53$).
This is most likely explained by the inaccurate initial imputations produced by \texttt{missForest}, resulting in the BART model being trained on incorrect values, compromising the overall model fit.

In the u-shape scenario, missBART1, missBART2, and BART\_cc perform similarly across all thresholds, implying little difference in the out-of-sample prediction and imputation accuracy, regardless of the detection probabilities.
In contrast, missBART2 dominates the other models in the n-shape scenario, while missBART1 has a slightly worse performance compared to BART\_cc.
Additionally, the RMSEs decrease as $p_t$ increases in Figure \ref{subfig: n-shape RMSE}, indicating that the models perform better on data which are more likely to be observed.
This is unsurprising, as the missing values lie further away from the observed values, making it exceptionally challenging for inappropriate models to capture the data's true extremes.

\section{Bivariate examples: additional details} 
\label{Appendix: bivariate examples}
Additional details are provided for the bivariate simulation studies reported in Section \ref{subsec: multivariate examples}, in the form of an outline of the simulation recipe in \ref{Appendix: bivariate simulation details} and additional results pertaining to RMSE and CRPS in \ref{Appendix: bivariate RMSE and CRPS}.

\subsection{Bivariate simulation details}
\label{Appendix: bivariate simulation details}
A summary of the simulation details is given in Table \ref{tab: Bivariate simulation recipe}.
\begin{table}[H]
    \centering
    \begin{adjustbox}{max width=\textwidth}
    \begin{tabular}{|c|c|c|c|c|c|}
        \hline
        & & \textbf{MAR 1} & \textbf{MAR 2} & \textbf{MNAR 1} & \textbf{MNAR 2} \\
        \hline
        \multirow{4}{*}{\textbf{Data Model}} & $n$ & \multicolumn{4}{c|}{$2000 \quad (n_{train} = 1500, n_{test} = 500)$} \\
        \cline{2-6}
        & $p$ & \multicolumn{4}{c|}{$2$} \\
        \cline{2-6}
        & $q$ & \multicolumn{4}{c|}{$5$} \\
        \cline{2-6}
        & \#Data trees & \multicolumn{4}{c|}{$8$} \\
        \hline
        \multirow{4}{*}{\textbf{Missingness Model}} & Model & Probit Regression & Probit BART & Probit Regression & Probit BART \\
        \cline{2-6}
        & Model covariates & X only & X only & Y only & Y only \\
        \cline{2-6}
        & \#Missing trees & - & 3 & - & 5 \\
        \cline{2-6}
        & Observed proportion & (57.10\%, 78.05\%) & (86.50\% 56.10\%) & (66.00\%, 50.45\%) & (58.05\%, 75.15\%) \\
        \hline
    \end{tabular}
    \end{adjustbox}
    \caption{Simulation recipes for the bivariate simulation studies. The complete dataset is consistent across all $4$ scenarios, while the missingness model follows a multivariate probit regression or probit BART model.}
    \label{tab: Bivariate simulation recipe}
\end{table}

\subsection{Out-of-sample RMSE and CRPS}
\label{Appendix: bivariate RMSE and CRPS}
To assess model performance and calibration, we compute and compare the out-of-sample RMSE and CRPS of the 8 models from Section \ref{subsec: bivariate examples}. 
We compute the RMSE and CRPS for each of the $p$ responses, separately for the missing, observed, and combined responses. This analysis thus shows more granular information about predictive performance than the Frobenius norms reported in Figure \ref{fig: bivariate Frobnorms}.

\begin{figure}[H]
    \centering
    \includegraphics[width=.99\linewidth]{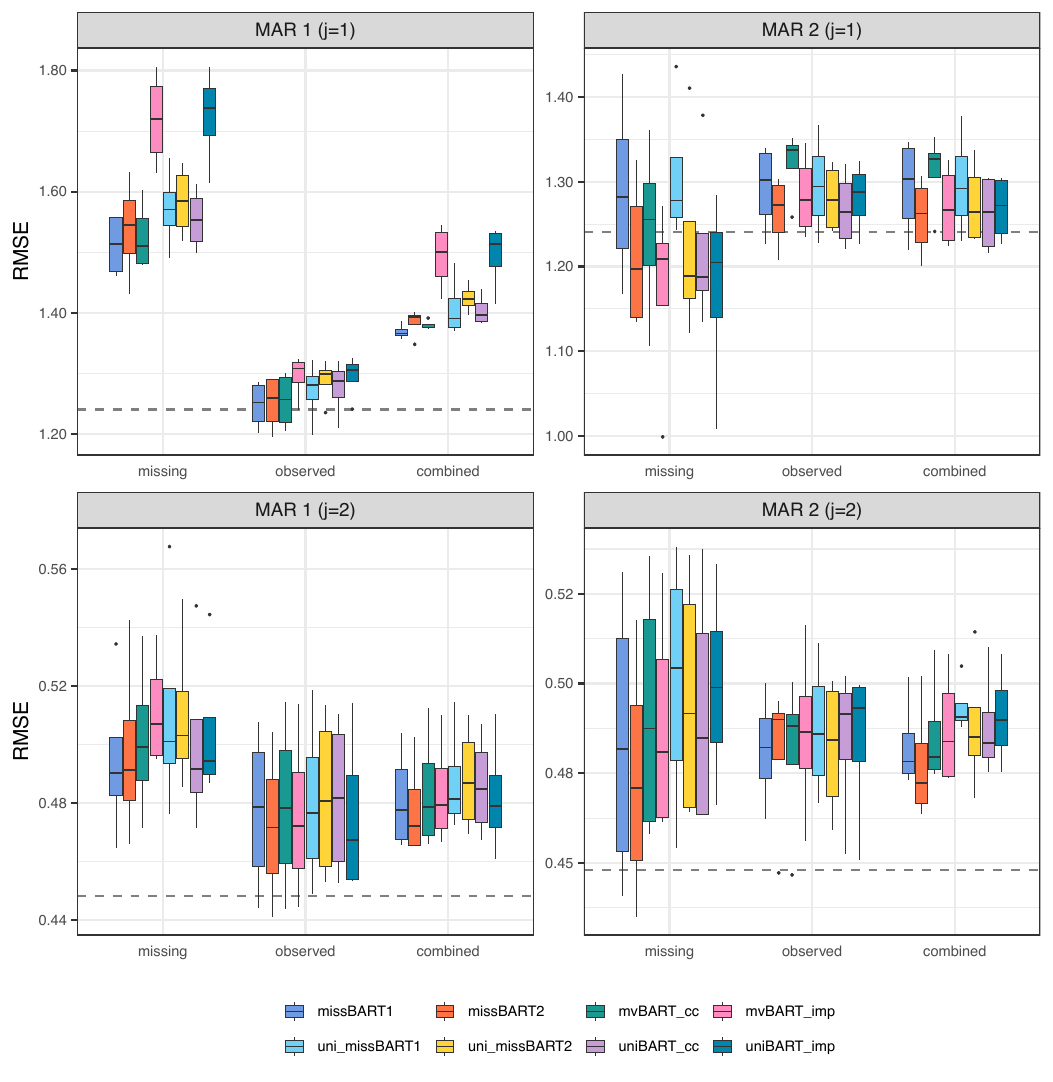}
    \caption{RMSEs for 8 different models in MAR 1 (left) and MAR 2 (right) scenarios, calculated across the missing, observed, and combined responses.}
    \label{fig: RMSE MAR}
\end{figure}

\begin{figure}[H]
    \centering
    \includegraphics[width=.99\linewidth]{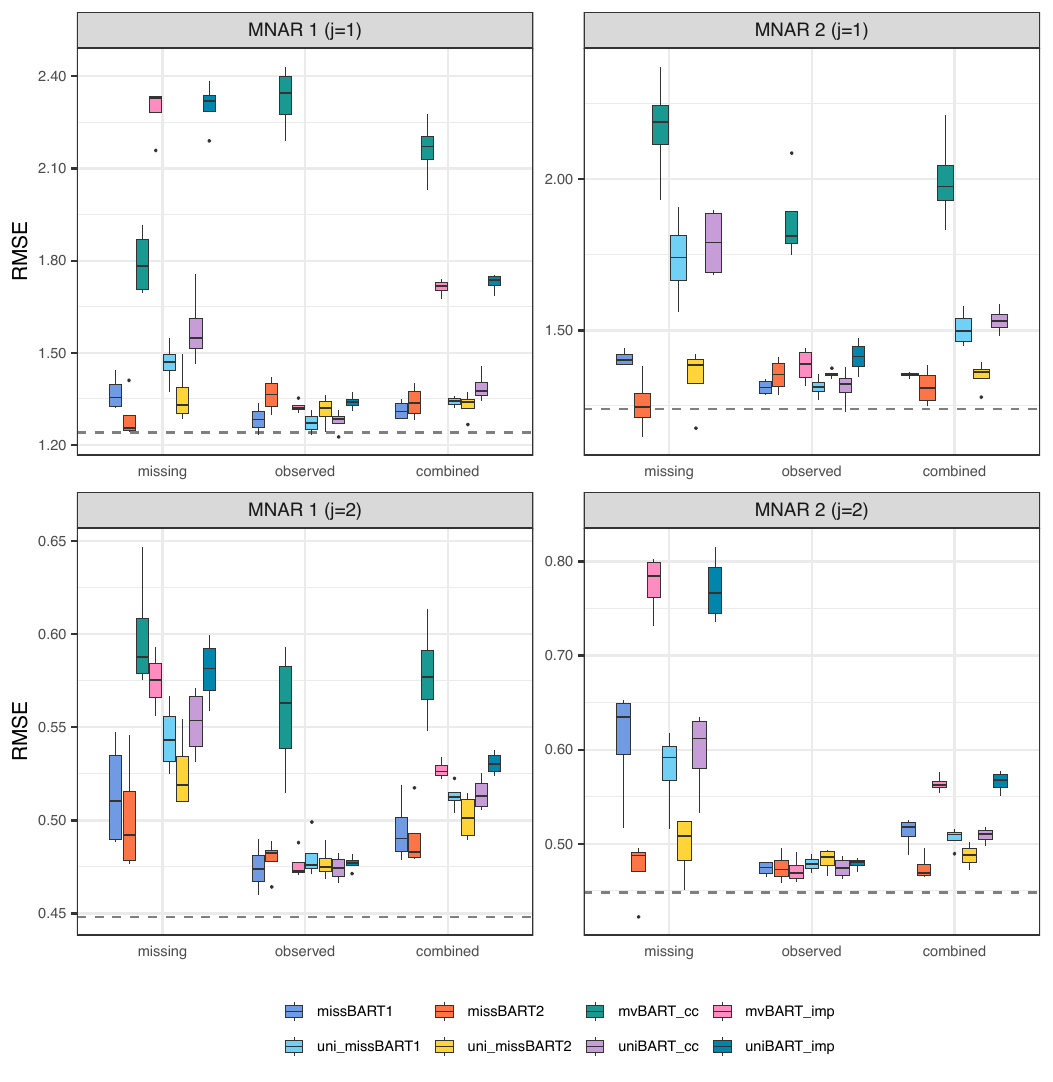}
    \caption{RMSEs for 8 different models in MNAR 1 (left) and MNAR 2 (right) scenarios, calculated across the missing, observed, and combined responses. Some results of mvBART\_cc, mvBART\_imp, and uniBART\_imp were omitted due to their poor performance in terms of RMSE.}
    \label{fig: RMSE MNAR}
\end{figure}

\begin{figure}[H]
    \centering
    \includegraphics[width=\linewidth]{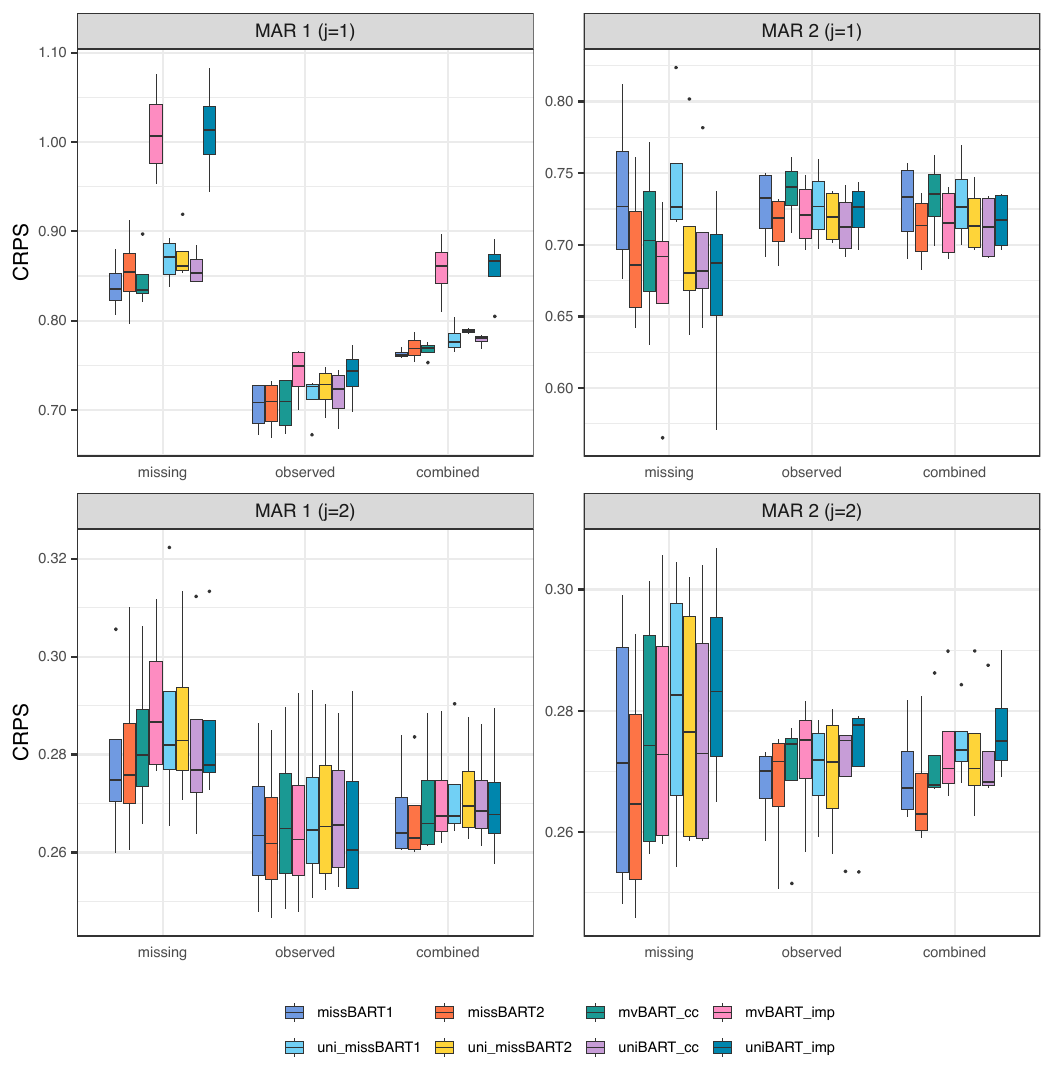}
    \caption{CRPS for 8 different models in MAR 1 (left) and MAR 2 (right) scenarios, calculated across the missing, observed, and combined responses.}
    \label{fig: CRPS MAR}
\end{figure}

\begin{figure}[H]
    \centering
    \includegraphics[width=\linewidth]{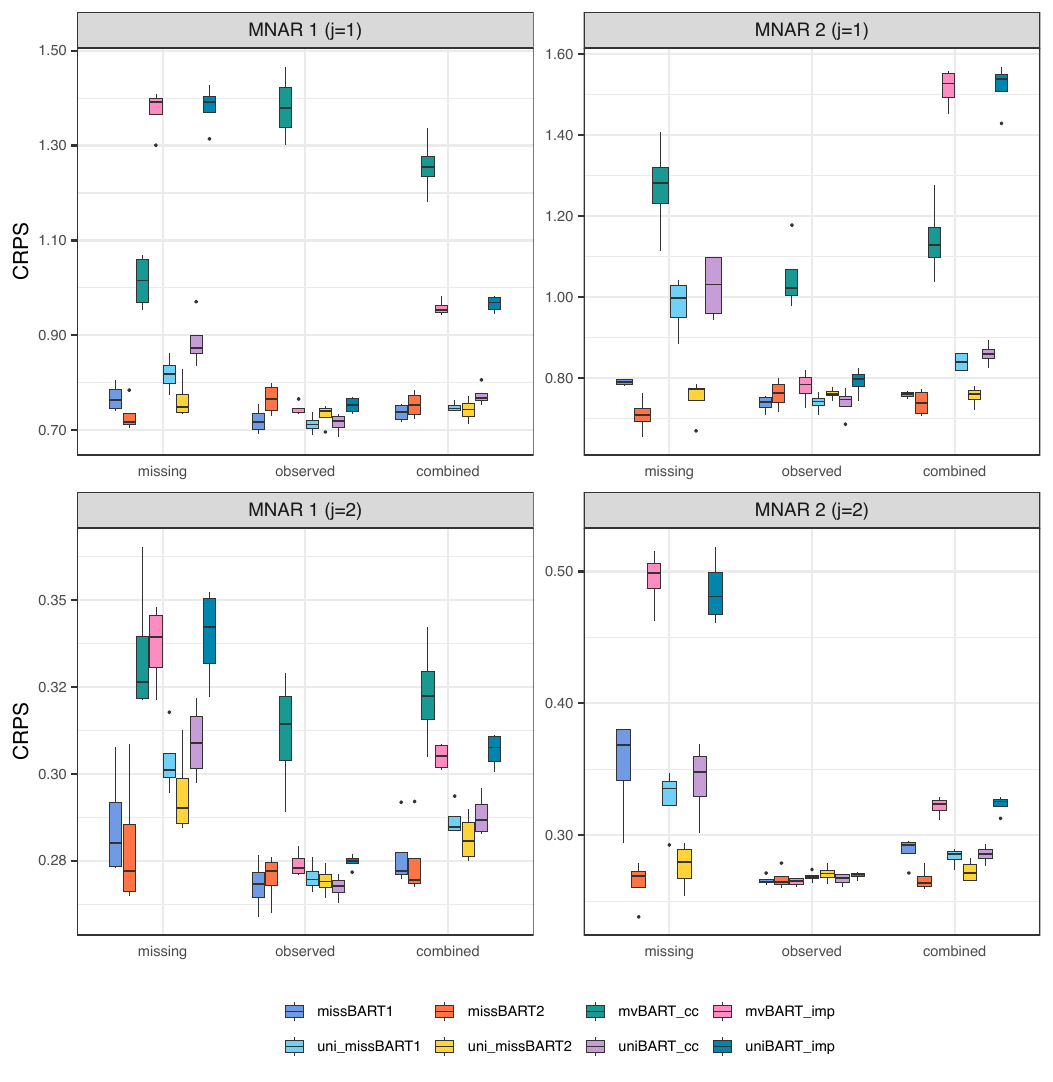}
    \caption{RMSEs for 8 different models in MNAR 1 (left) and MNAR 2 (right) scenarios, calculated across the missing, observed, and combined responses. Some results of mvBART\_cc, mvBART\_imp, and uniBART\_imp were omitted due to their poor performance in terms of RMSE.}
    \label{fig: CRPS MNAR}
\end{figure}

Aligning with the results from Section \ref{subsec: bivariate examples}, the joint models perform at least as well as the complete-case models for both missing and observed responses under MAR missingness, and perform better than other models under MNAR, as complete-case and \texttt{missForest}-imputed models fail to account for any relationship between the responses and their corresponding missingness status.
These are shown in Figures \ref{fig: RMSE MAR} and \ref{fig: RMSE MNAR} for the RMSE and Figures \ref{fig: CRPS MAR} and \ref{fig: CRPS MNAR} for the CRPS.
In Figure \ref{fig: RMSE MNAR} and Figure \ref{fig: CRPS MNAR}, some results of mvBART\_cc, mvBART\_imp, and uniBART\_imp were omitted due to overly high RMSE or CRPS values. 

The \texttt{missForest} imputed models show poor performance across MAR 1, MNAR 1, and MNAR 2, which can again be attributed to the inaccurate imputations obtained prior to model fitting leading to erroneous results. 
Overall, missBART2 demonstrated superior performances in both the MAR and MNAR scenarios, particularly in making accurate posterior imputations for the missing responses. 


\section{Multivariate examples: additional details}
We provide additional details and results for the multivariate simulated scenarios with missing covariates in Section \ref{subsec: multivariate examples}, namely MNAR\_amp1 and MNAR\_amp2.
In \ref{Appendix: missing X patterns}, we include the missingness patterns induced in the covariates for both scenarios.
The out-of-sample RMSEs are shown in \ref{Appendix: MNAR_amp RMSE}.

\subsection{Covariate missing patterns}
\label{Appendix: missing X patterns}
The missingness patterns induced in the covariates under MNAR\_amp1 and MNAR\_amp2 are shown below:
\begin{figure}[H]
\centering
\begin{minipage}{.5\textwidth}
  \centering
  \includegraphics[width=\linewidth]{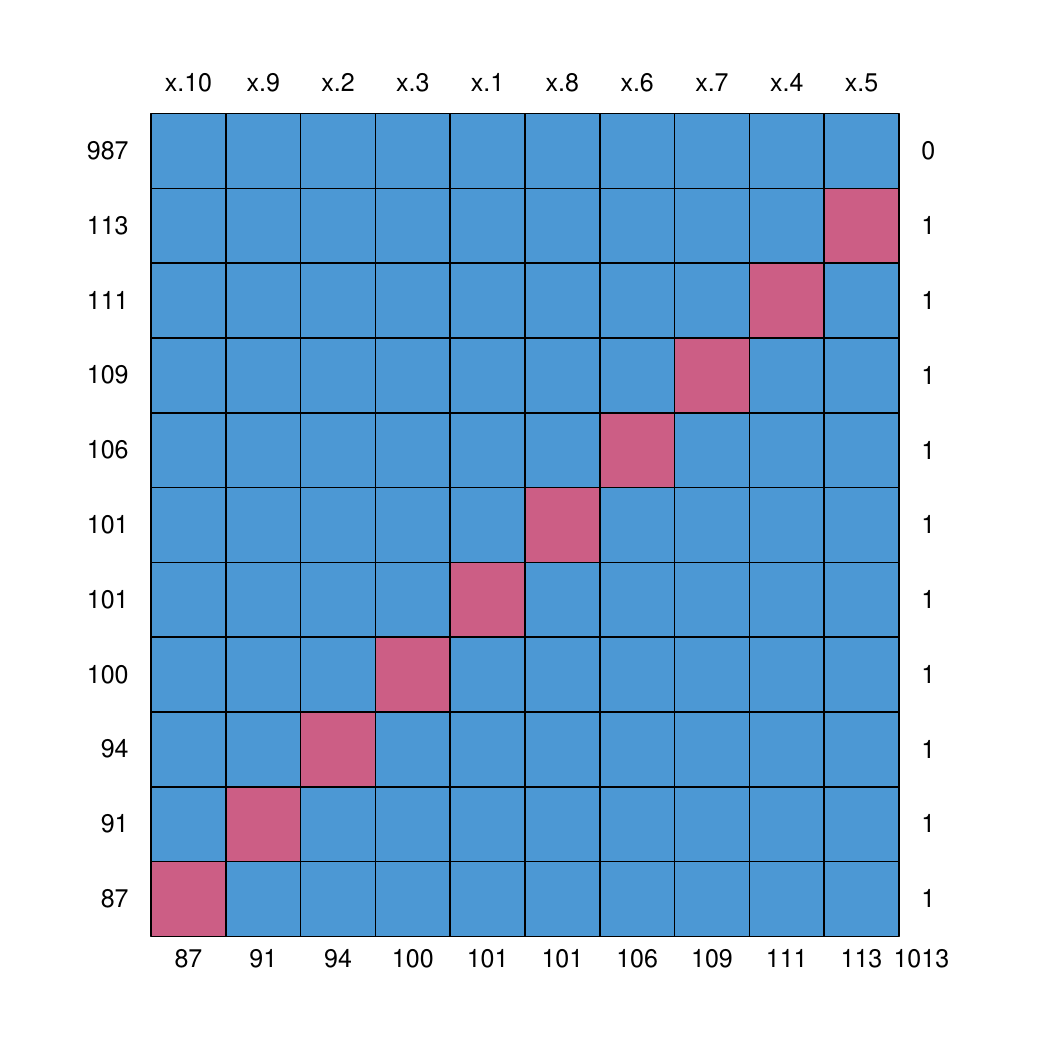}
  \captionof{figure}{}
  \label{Appendix: missx1}
\end{minipage}%
\begin{minipage}{.5\textwidth}
  \centering
  \includegraphics[width=\linewidth]{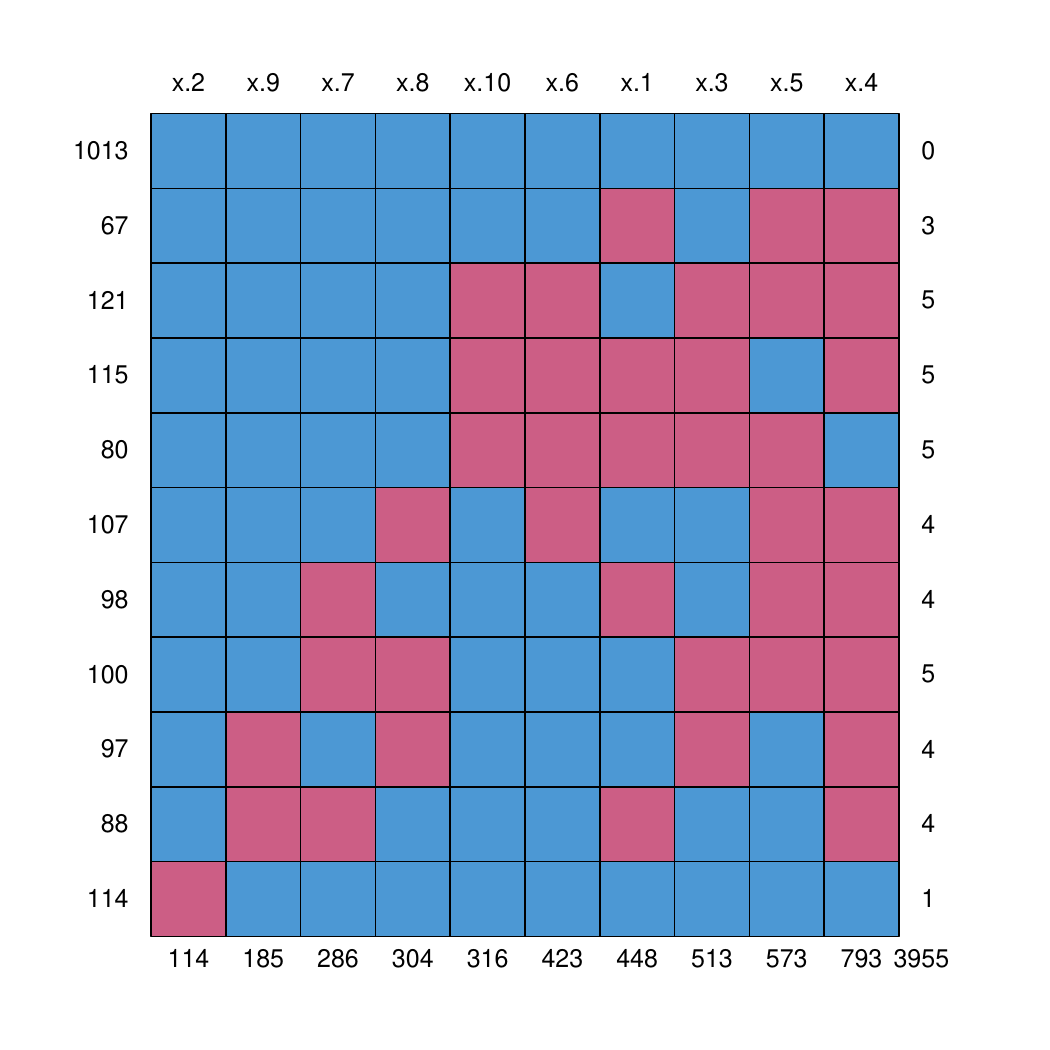}
  \captionof{figure}{}
  \label{Appendix: missx2}
\end{minipage}
\end{figure}

\subsection{Out-of-sample RMSE}
\label{Appendix: MNAR_amp RMSE}
\begin{figure}[H]
    \centering
    \includegraphics[height=\dimexpr\textheight - \baselineskip\relax]{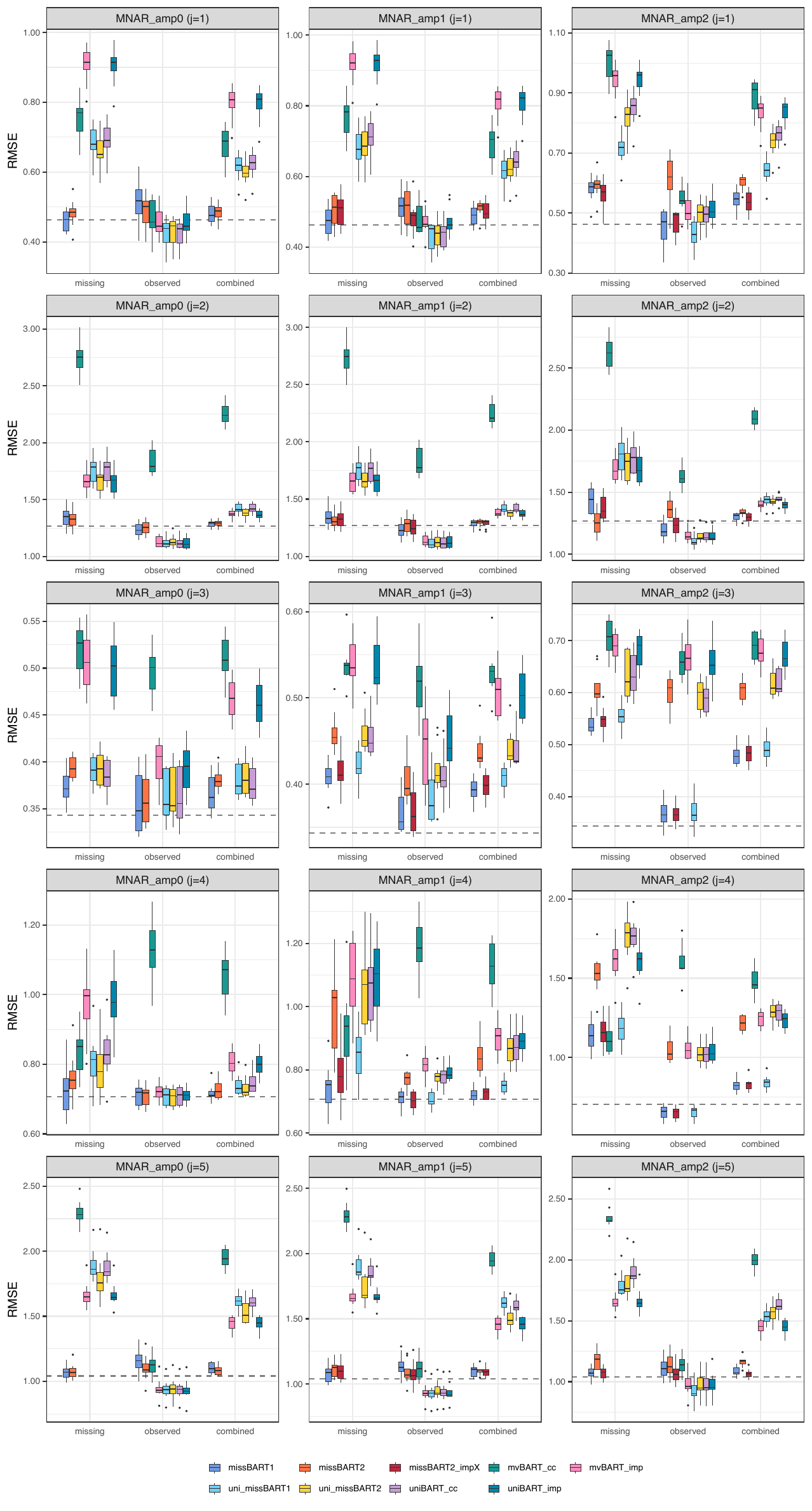}
    \caption{RMSEs for multivariate simulated scenarios with missingness in the covariates.}
    \label{fig: RMSE MNAR_amp}
\end{figure}

The out-of-sample RMSEs for MNAR\_amp0, MNAR\_amp1, and MNAR\_amp2 are shown in Figure \ref{fig: RMSE MNAR_amp}. Similar to \ref{Appendix: bivariate RMSE and CRPS}, RMSEs are shown separately for the missing, observed, and combined univariate responses.

\section{\textit{global Amax}: additional details}
\label{Appendix: global amax}
We give additional details and results from Section \ref{sec: 7 Real data example} in \ref{Appendix: missing patterns response} and \ref{Appendix: PDP Miq Pavail}, which show the missingness patterns of the responses and additional PDP and ICE plots, respectively.

\subsection{Missingness patterns in response variables}
\label{Appendix: missing patterns response}
\begin{figure}[H]
    \centering
    \includegraphics[width=0.5\textwidth, trim={0 2cm 0 2.9cm}, clip]{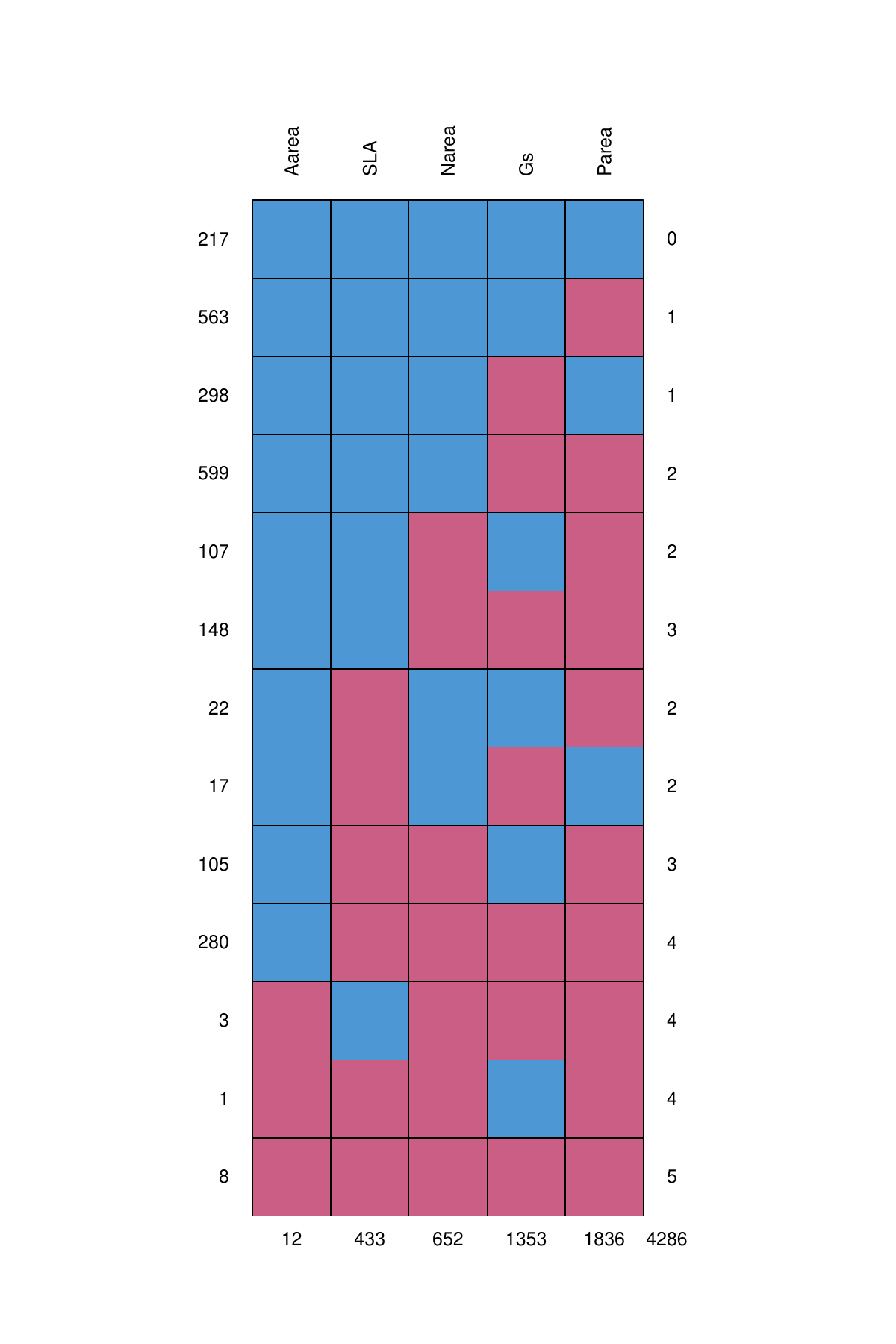}
    \caption{Missingness patterns in the \textit{global Amax} responses. Blue indicates observed responses, red indicates missing responses. Top column labels show the response variables. Left row labels show the number of cases with the unique missingness patterns. Right row labels denote the total number of missing variables within that pattern. Bottom column labels show the number of cases where each variable is missing.}
    \label{fig: plant_missing_pattern}
\end{figure}

The missingness patterns for the response variables of the \textit{global Amax} data are shown in Figure \ref{fig: plant_missing_pattern}. Recall that the covariates in the \textit{global Amax} data are fully observed.

\subsection{PDP and ICE plots for \textit{Miq} and \textit{Pavail}}
\label{Appendix: PDP Miq Pavail}
\begin{figure}[H]
    \begin{subfigure}[b]{\textwidth}
        \centering
        \includegraphics[width=\textwidth, trim={0 0.2cm 0 0}, clip]{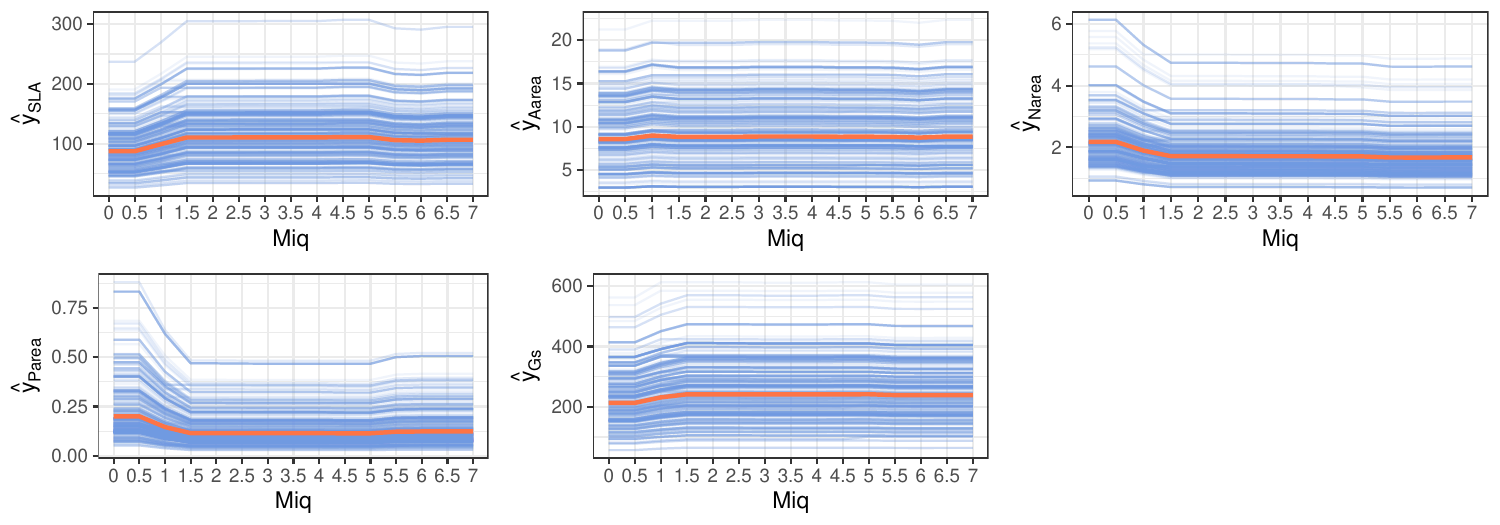}
        \caption{PDP + ICE curves for all responses across different levels of \textit{Miq}.}
        \label{subfig: Miq pdp}
    \end{subfigure}
    \vfill
    \begin{subfigure}[b]{\textwidth}
        \centering
        \includegraphics[width=\textwidth, trim={0 0.2cm 0 0}, clip]{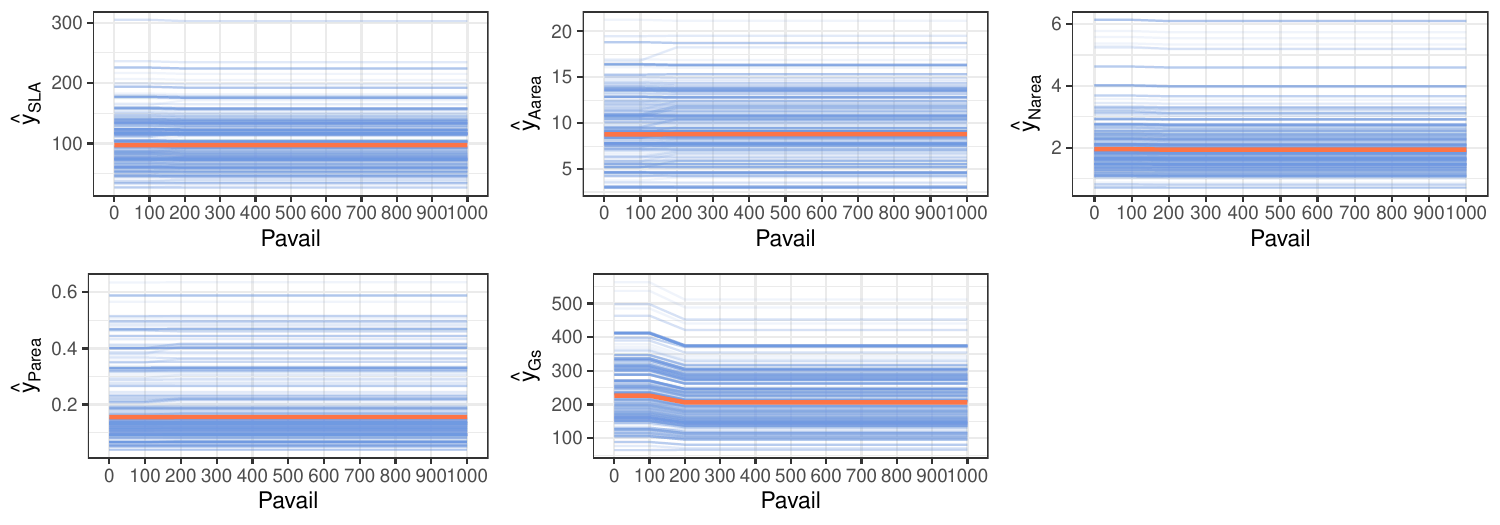}
        \caption{PDP + ICE curves for responses across different levels of \textit{Pavail}.}
        \label{subfig: Pavail pdp}
    \end{subfigure}
    \caption{PDP + ICE plots for all responses across \textit{Miq} and \textit{Pavail} levels, two other variables deemed as influential variables by \citet{maire2015global}.}
    \label{fig: PDP Miq Pavail}
\end{figure}

Figure \ref{fig: PDP Miq Pavail} shows two additional PDP and ICE plots from the missBART2 regression trees.
\citet{maire2015global} found that \textit{SLA} increases while \textit{Aarea}, \textit{Narea}, and \textit{Parea} decrease as \textit{Miq} increases, and only \textit{Parea} increases while \textit{Gs} decreases with increasing \textit{Pavail}.
From Figure \ref{subfig: Miq pdp}, \textit{SLA} and \textit{Gs} increase, \textit{Aarea} remains unchanged, and \textit{Narea} and \textit{Parea} decrease as \textit{Miq} increases. 
Apart from a slight decrease in \textit{Gs}, Figure \ref{subfig: Pavail pdp} shows virtually no changes in all other responses as \textit{Pavail} increases.

\end{document}